\documentclass[preprintnumbers,amsmath,amssymb]{revtex4}
\usepackage{graphicx}
\usepackage{dcolumn}% Align table columns on decimal point
\usepackage{bm}% bold math
\usepackage{graphicx}
\usepackage{epsfig}
\usepackage[usenames]{color}

\usepackage{setspace}
\usepackage[mathscr]{eucal}
\usepackage{subfig}
\usepackage{amsthm}
\usepackage{hyperref}
\usepackage{float}
\usepackage[utf8]{inputenc}  %for apostrophe

\begin{document}

\title{Pulses and Snakes in Ginzburg--Landau Equation}

\author{Stefan C. Mancas}
\email{mancass@erau.edu}
\affiliation{Department of Mathematics, Embry-Riddle Aeronautical University,\\ Daytona-Beach, FL. 32114-3900, U.S.A.}

\author{Roy S. Choudhury}
\affiliation{Department of Mathematics, University of Central Florida,\\ Orlando, FL. 32816-1364, U.S.A}
\email{choudhur@longwood.cs.ucf.edu}

\begin{abstract}

Using a variational formulation for partial differential equations (PDEs) combined with numerical simulations on ordinary differential equations (ODEs), we find two categories (pulses and snakes) of dissipative solitons, and analyze the dependence of both their shape and stability on the physical parameters of the cubic-quintic Ginzburg-Landau equation (CGLE).  In contrast to the regular solitary waves investigated in numerous integrable and non-integrable systems over the last three decades, these dissipative solitons are not stationary in time. Rather, they are spatially confined pulse-type structures whose envelopes exhibit complicated temporal dynamics. Numerical simulations  reveal very interesting bifurcations sequences  as the parameters of the CGLE are varied. Our predictions on the variation of the soliton amplitude, width, position, speed and phase of the solutions using the variational formulation agree with simulation results.

First, we develop a variational formalism  which explores the various classes of dissipative solitons. Given the complex dynamics  the trial functions have been generalized considerably over conventional ones to keep the shape relatively simple, and the trial function integrable while allowing arbitrary temporal variation of the amplitude, width, position, speed and phase of the pulses and snakes. 

In addition, the resulting Euler-Lagrange (EL) equations from the variational formulation are treated in a completely novel way. Rather than consider the stable fixed points which correspond to the well-known stationary solitons, we use dynamical systems theory to focus on more complex attractors viz. periodic (pulses) and  quasiperiodic (snakes). Periodic evolution of the trial function parameters on stable periodic attractors yield solitons whose amplitudes and widths are non-stationary or time dependent. 

Secondly, we investigate the dissipative solitons of the CGLE and analyze its qualitative behavior by using numerical methods for ODEs. To solve numerically the nonlinear systems of ODEs that represent EL equations obtained from variational technique, we use an explicit Runge-Kutta fourth order method (RK4).

Finally, we elucidate the Hopf bifurcation mechanism responsible for the various pulsating solitary waves, as well as its absence in Hamiltonian and integrable systems where such structures are absent due to the lack of dissipation.
\end{abstract}

\maketitle

\section{Introduction}\label{sec:1}

The cubic-quintic  Ginzburg-Landau equation is the canonical equation governing the weakly nonlinear behavior of dissipative systems in a wide variety of disciplines \cite{Dodd}. In fluid mechanics, it is also often referred to as the Newell-Whitehead equation after the authors who derived it in the context of B\'enard convection \cite{Dodd,Drazin:1}.

Many basic properties of the equation and its solutions are reviewed in \cite{Aranson,Bowman,Saarloos}, together with applications to a vast variety of phenomena including nonlinear waves, second-order phase transitions, superconductivity, superfluidity, Bose-Einstein condensation, liquid crystals and string theory. Numerical studies by Brusch {\it et. al.} \cite{Brusch:2,Brusch:1} which primarily consider periodic traveling wave solutions of the cubic CGLE, together with secondary pitchfork bifurcations and period doubling cascades into disordered turbulent regimes, also give comprehensive summaries of other work on this system. Early numerical studies \cite{Keefe,Landman} and theoretical investigations \cite{Newton:1,Newton:2} of periodic solutions and secondary bifurcations are also of general interest for our work here. This has proved to be a rich system with very diverse solution behaviors. In particular, a relatively early and influential review by van Saarloos and Hohenberg \cite{Saarloos}, also recently extended to two coupled cubic CGLE equations \cite{Hecke,Alvarez}, considered phase-plane counting arguments for traveling wave coherent structures, analytic and perturbative solutions, limited comparisons to numerics, and so-called ``linear marginal stability analysis" to select the phase speed of the traveling waves.

In this article we will only refer to two sets of studies which will directly pertain to our work.  The first class of papers \cite{Holmes,Doelman:1,Doelman:2,Duan,Doelman:3} used dynamical systems techniques to prove that CGLE admits periodic and quasi-periodic traveling wave solutions, while the second class of papers \cite{Artigas,Brusch:1,Brusch:2}, primarily involving numerical simulations of the full CGLE  in the context of nonlinear optics, revealed various branches of plane wave solutions which are referred to as continuous wave (CW) solutions. More importantly, these latter studies also found various spatially confined coherent structures with envelopes which exhibit complicated temporal dynamics \cite{Akhmediev:1,Soto,Mancas:2,Soto:4}. All indications are that these classes of solutions, all of which have amplitudes that vary in time, do not exist as stable structures in Hamiltonian systems. Even if excited initially, amplitude modulated solitons restructure into regular stationary solutions. Exceptions to this rule are the integrable models where the pulsating structures are nonlinear superpositions or fundamental solutions \cite{Satsuma}. Therefore, these classes of solutions are novel and they exist only in the presence of dissipation.

In this context, we note that numerous attempts have been made to extend the well-developed concept of soliton interactions in integrable, conservative systems \cite{Drazin:2,Nody3,Nody4} to more realistic active or dissipative media which are governed by non-integrable model equations. The reason is that the complicated spatio-temporal dynamics of coherent structures are governed by systems of ODEs, or low-dimensional dynamical systems, rather than by the original complex nonlinear PDE model. Hence, various theoretical approaches may be brought to bear on these ODEs. This is appropriate, particularly where the dynamics of dissipative systems is primarily governed by localized coherent structures such as pulses (solitary waves) and kinks (fronts or shocks)  \cite{Saarloos,Drazin:2,Murray,Balmforth}. Since these structures correspond to spatial modulations, they are also often referred to  spatially-localized ``patterns", and they may be information carriers such as in optics. The speeds and locations of them may vary in a complex manner as they interact, but their spatial coherence is preserved. Coherent structures may be transitory when they are unstable to small disturbances in some neighborhood of the existence.

In the language of the Los Alamos school, the fully spatiotemporal approach \cite{Nody3} followed here may be said to be the ``collective coordinates" formulation. In other words, we consider a pulse or solitary wave at any time as a coherent collective entity (or coordinate). This solitary wave is then temporally modulated. The main spatial approach proposed, and explored, in this paper is the \underline{variational method}, while the temporal approach is based on solving \underline{numerically} the complicated system of ODEs resulting from the EL equations. We would  like to particularly cite Kaup and Malomed\rq{}s work \cite{Kaup:1,Kaup:2,Kaup:3,Kaup:4} in constructing regular and embedded solitons of various complicated $\chi^\mathrm{2}-\chi^\mathrm{3}$ systems. These were instrumental in focusing our attention on this method, and attempting to extend its use to new classes of dissipative solitons. 

Given this setting, Section~\ref{sec:2} outlines the generalized variational formulation including  novel  trial functions to be employed in modeling the pulsating and snaking solitary waves.   In  Section~\ref{sec:3} we firstly elucidate the new mechanism responsible for the various classes of pulsating solitary wave solutions in dissipative systems, viz. the possibility of Hopf bifurcations. This also explains the absence of pulsating and snakes solitary waves in Hamiltonian and integrable systems.  Periodic evolution of the trial function parameters on stable periodic attractors resulting from supercritical Hopf bifurcations, when substituted back into the trial function, yield pulsating solitary waves. Fourth order Runge-Kutta (RK4) method  \cite{Williamson} is used to solve numerically he nonlinear EL equations. In  Section~\ref{sec:4} we  developed a generalized  multiple scales analysis  to construct analytical approximations for the periodic orbits arising through Hopf bifurcation of the fixed point of the EL equations ~(\ref{4.5}) or ~(\ref{5.5}).  Sections~\ref{sec:5} and ~\ref{sec:6}  provide numerical results for the pulsating and snakes solitons. Section ~\ref{sec:8} summarizes the results. 

\section{The Generalized Variational Formulation}\label{sec:2}

We develop a general variational formulation  \cite{Nody2,Kaup:1,Kaup:2} to address the pulses on all parameter ranges. As mentioned earlier, we shall need to generalize previous variational approaches in several crucial ways.

We shall consider the CGLE in the form \cite{Saarloos}
\begin{equation}\label{2.1}
\partial_tA=\epsilon A+(b_1+ic_1)\partial_x^2A-(b_3-ic_3) |A|^2A-(b_5-ic_5) |A|^4 A,
\end{equation}
in which we chose a frame without the linear $A_x$ term, and without imaginary linear term $i c_0 A$, which implies that all parameters are real, including the gain/loss coefficient $\epsilon$. Note that any of the  three of the coefficients (no two of which are in the same term) may be set to unity by appropriate scalings of time, space and $A(x,t)$, but that will possibly cause a loss in physical interpretation of the parameters of the system. 

The interpretation of the system's parameters in (\ref{2.1}) depends on the particular field of work. In optics, $A(x;t)$ is the normalized envelope of the field, $x$ is the transversal coordinate, $t$ is the propagation distance or the cavity number.  The system's parameters of (\ref{2.1}) are: $\epsilon$ linear loss/gain, $b_1$- angular spectral filtering, $c_1=0.5$- second-order diffraction coefficient, $b_3$- nonlinear gain/loss, $c_3=1$- nonlinear dispersion, $b_5$- saturation of the nonlinear gain/loss, and $c_5$- saturation of the nonlinear refractive index. 

Proceeding as in \cite{Kaup:1}, the pseudo-Lagrangian (Lagrangian for dissipative systems) of ~(\ref{2.1}) may be written as
\begin{align}\label{3.1}
\mathscr{L}&=r^*\big[\partial_tA-\epsilon A-(b_1+ic_1)\partial_x^2A+(b_3-ic_3) |A|^2A+(b_5-ic_5) |A|^4 A\big]\notag\\
&+r\big[\partial_tA^*-\epsilon A^*-(b_1-ic_1)\partial_x^2A^*+(b_3+ic_3) |A|^2A^*+(b_5+ic_5) |A|^4 A^*\big]
\end{align}
Here $r$ is the usual variable which is the  auxiliary field employed in \cite{Kaup:1} and it satisfies a perturbative evolution equation dual to the CGLE with all non-Hamiltonian terms reversed in sign.

The second key assumption involves the trial functions $A(t)$ and $r(t)$ which have been generalized considerably over conventional ones to keep the shape relatively simple and the trial functions integrable.
To this end, we choose Gaussian ansatz of the form:
\begin{align}\label{3.2}
A(x,t)&=A_1(t)e^{-\sigma_1(t)^2\lbrack x-\phi_1(t) \rbrack^2}e^{i\alpha_1(t)}\\
r(x,t)&=e^{-\sigma_2(t)^2\lbrack x-\phi_2(t) \rbrack^2}e^{i\alpha_2(t)}\notag
\end{align}
where $A_1(t)$ is the amplitude, $\sigma_i(t)$'s are the inverse widths, $\phi_i(t)$'s are the positions (with $\dot{\phi}_i(t)$ the phase  speed), and $\alpha_i(t)$'s are the phases of the solitons, and  are all allowed to vary arbitrarily in time. For now, the chirp terms $e^{i\beta(x)} $ are omitted for simplicity. Substituting ~(\ref{3.2})  in ~(\ref{3.1}) the effective or averaged Lagrangian is
\begin{eqnarray}\label{3.4}
&&L_{EFF}=\int_{-\infty}^{\infty}\mathscr{L}dx=2\sqrt{\pi}\Bigg\{-\frac{e^{-\frac{\sigma_1(t)^2\sigma_2(t)^2[\phi_1(t)-\phi_2(t)]^2}{\sigma_1(t)^2+\sigma_2(t)^2}}}{[\sigma_1(t)^2+\sigma_2(t)^2]^\frac12}\epsilon A_1(t)\cos[\alpha_1(t)-\alpha_2(t)] \nonumber \\
&&+\frac{e^{-\frac{3\sigma_1(t)^2\sigma_2(t)^2[\phi_1(t)-\phi_2(t)]^2}{3\sigma_1(t)^2+\sigma_2(t)^2}} }{\Big[3\sigma_1(t)^2+\sigma_2(t)^2\Big]^\frac12}A_1(t)^3\Bigg[b_3\cos[\alpha_1(t)-\alpha_2(t)]+c_3\sin[\alpha_1(t)-\alpha_2(t)]\Bigg] \nonumber \\
&&+\frac{e^{-\frac{5\sigma_1(t)^2\sigma_2(t)^2[\phi_1(t)-\phi_2(t)]^2}{5\sigma_1(t)^2+\sigma_2(t)^2}} }{\Big[5\sigma_1(t)^2+\sigma_2(t)^2\Big]^\frac12}A_1(t)^5\Bigg[b_5\cos[\alpha_1(t)-\alpha_2(t)]+c_5\sin[\alpha_1(t)-\alpha_2(t)]\Bigg] \nonumber\\
&&+\frac{e^{-\frac{\sigma_1(t)^2\sigma_2(t)^2[\phi_1(t)-\phi_2(t)]^2}{\sigma_1(t)^2+\sigma_2(t)^2}}}{\Big[\sigma_1(t)^2+\sigma_2(t)^2\Big]^\frac52}\Bigg[\cos[\alpha_1(t)-\alpha_2(t)][\sigma_1(t)^2+\sigma_2(t)^2]^2\dot{A}_1(t)\nonumber
\end{eqnarray}
\begin{eqnarray}
&&+A_1(t)\Bigg(-2\sigma_1(t)^2\sigma_2(t)^2\Big[b_1\cos[\alpha_1(t)-\alpha_2(t)]-c_1\sin[\alpha_1(t)-\alpha_2(t)]\Big]\Big[-\sigma_2(t)^2 \nonumber \\
&&+\sigma_1(t)^2[-1+2\sigma_2(t)^2[\phi_1(t)-\phi_2(t)]^2]\Big]-\dot{\alpha_1}(t)\sin[\alpha_1(t)-\alpha_2(t)][\sigma_1(t)^2+\sigma_2(t)^2]^2\nonumber \\
&&-\sigma_1(t)\dot{\sigma_1}(t)\cos[\alpha_1(t)-\alpha_2(t)]\Big[\sigma_1(t)^2+\sigma_2(t)^2+2\sigma_2(t)^4[\phi_1(t)-\phi_2(t)]^2\Big] \nonumber \\
&&-2\dot{\phi_1}(t)\sigma_1(t)^2\sigma_2(t)^2[\phi_1(t)-\phi_2(t)][\sigma_1(t)^2+\sigma_2(t)^2]\cos[\alpha_1(t)-\alpha_2(t)]\Bigg)\Bigg]\Bigg\}
\end{eqnarray}

Since ~(\ref{3.4}) reveals that only the relative phase 
$ \alpha(t)=\alpha_1(t)-\alpha_2(t)$ of $A(x,t)$ and $r(x,t)$ is relevant, we henceforth take
\begin{align}
&\alpha_1(t)=\alpha(t)\notag\\
&\alpha_2(t)=0 \label{3.5}
\end{align}
with no loss of generality. 
Also, for algebraic tractability, we have found it necessary to assume 
\begin{equation}\label{3.6}
\sigma_2(t)=m\sigma_1(t)\equiv m\sigma(t).
\end{equation}
While this ties the widths of the $A(x,t)$ and $r(x,t)$ fields together, the loss of generality is acceptable since the field $r(x,t)$ has no real physical significance.
For reasons of algebraic simplicity, we may also scale the positions according to:
\begin{align}
&\phi_1(t)=\phi(t) \notag\\ 
&\phi_2(t)=0, \label{3.7}
\end{align}
although this assumption may easily be relaxed. In fact, we may expect that it may be necessary to relax ~(\ref{3.7}) for certain classes of dissipative solitons.
 
Hence, using all assumptions, (i.e. ~(\ref{3.5})-~(\ref{3.7}) in ~(\ref{3.4})), the effective Lagrangian ~(\ref{3.4}) may be written in a simpler but still general form
\begin{eqnarray}\label{3.9}
L_{EFF} & = & 2\sqrt{\pi}\Bigg\{\frac{A_1(t)}{\sigma(t)}\Bigg[-\frac{e^{-\frac{m^2\sigma(t)^2\phi(t)^2}{1+m^2}}}{[1+m^2]^\frac12}\epsilon \cos\alpha(t) \nonumber \\
 &&  \qquad \qquad \qquad + \frac{e^{-\frac{3m^2\sigma(t)^2\phi(t)^2}{3+m^2}}}{[3+m^2]^\frac12}A_1(t)^2\Big[b_3\cos\alpha(t)+c_3\sin\alpha(t)\Big] \nonumber \\
 &&  \qquad \qquad \qquad + \frac{e^{-\frac{5m^2\sigma(t)^2\phi(t)^2}{5+m^2}}}{[5+m^2]^\frac12}A_1(t)^4\Big[b_5\cos\alpha(t)+c_5\sin\alpha(t)\Big]\Bigg] \nonumber\\
 &&  \qquad \qquad \qquad + \frac{e^{-\frac{m^2\sigma(t)^2\phi(t)^2}{1+m^2}}}{[1+m^2]^\frac52\sigma(t)^2}\Bigg[(1+m^2)^2\cos\alpha(t)\sigma(t)\dot{A_1}(t) \nonumber\\
 && \qquad \qquad \qquad -
A_1(t)\Bigg(4m^4\sigma(t)^5\phi(t)^2\Big[b_1\cos\alpha(t)-c_1\sin\alpha(t)\Big] \nonumber \\ 
&& \qquad \qquad \qquad +(1+m^2)^2\sigma(t)\dot{\alpha}(t)\sin\alpha(t)+(1+m^2)\dot{\sigma}(t)\cos\alpha(t)\nonumber \\
&& \qquad \qquad \qquad - 2m^2(1+m^2)\sigma(t)^3\Big[b_1\cos\alpha(t)-c_1\sin\alpha(t)\Big] \nonumber \\
&& \qquad \qquad \qquad + 2m^4\dot{\sigma}(t)\sigma(t)^2\phi(t)^2+\dot{\phi}(t)\phi(t)\cos\alpha(t)\Bigg)\Bigg]\Bigg\}
\end{eqnarray}

\section{Framework for Investigation of Euler-Lagrange Equations for Pulsating and Snake Solitons}\label{sec:3}

It is widely reported \cite{Artigas,Akhmediev:3} and generally accepted that Hamiltonian systems, as well as integrable systems which are a subclass, do not admit pulsating solitary wave solutions. If excited initially, pulsating solitons in Hamiltonian and integrable systems re-shape themselves and evolve into regular stationary waves. The only exceptions are pulsating structures comprising nonlinear superpositions of stationary solitons in integrable systems \cite{Satsuma}.

In addition, the regimes of the pulsating solitons in the CGLE are very far from the integrable nonlinear Schr\"odinger equation limit. This fact, and the great diversity of pulsating solitons in the CGLE, both indicate a new mechanism which is operative in dissipative systems in the creation of these pulsating structures.

The primary point of this paper is that Hopf bifurcations are the new mechanism responsible for the occurrence of these pulsating solitons in dissipative systems, and we shall analyze both plain pulsating solitons and snakes via this mechanism. However, in order to establish that Hopf bifurcations are indeed the operative mechanism creating the various pulsating solitons in dissipative systems, we first proceed to prove their absence in Hamiltonian systems. This will also explain the above-mentioned absence of pulsating solitons in Hamiltonian and integrable systems.

For a Hamiltonian system with Hamiltonian $H$, the particular evolution equations  may be represented  in canonical form as \cite{Fadeev}.
\begin{align}
&i \Psi_\zeta=\frac{\delta H}{\delta \Psi^{\star}}\notag\\
&i \Psi^{\star}_\zeta=-\frac{\delta H}{\delta \Psi}.\label{2.2}
\end{align}
These may be further combined  into
\begin{equation}
i \dot{\vec{x}}=L \nabla_{\vec{x}}H(\vec{x})\label{2.3}
\end{equation}
where $\dot{}$ denotes $\delta / \delta \zeta$,
\begin{equation}
\vec{x}=[\Psi,\Psi^{\star}],\label{2.4}
\end{equation}
$I$ is the $n\times n$ unit matrix, and $L$ is the symplectic gradient of $H(\vec{x})$
\begin{eqnarray}
L=
\left(\begin{array}{ccc}
0 & I\\
-I & 0 \\
\end{array}\right).\label{2.5}
\end{eqnarray}
Equation \eqref{2.3} follows from
\begin{eqnarray}
i \left(\begin{array}{ccc}
\dot{\Psi}\\
\dot{\Psi}^{\star} \\
\end{array}\right)=
\left(\begin{array}{ccc}
0 & I\\
-I & 0 \\
\end{array}\right) \left(\begin{array}{ccc}
\nabla_{\Psi}H\\
\nabla_{\Psi^{\star}}H\\
\end{array}\right)\notag
\end{eqnarray}
which is identical to \eqref{2.2}.

The fixed (or equilibrium or critical ) points of \eqref{2.3} satisfy
\begin{equation}
\nabla_{\vec{x}}H(\vec{x})=0, \label{2.6}
\end{equation}
or equivalently 

\begin{equation}
\frac{\delta H}{\delta \Psi^{\star}}=0, \qquad \frac{\delta H}{\delta \Psi}=0.\notag\\
\end{equation}
Using the standard representation
\begin{equation}
H=\frac 12 \langle \Psi_{\zeta},\Psi_{\zeta} \rangle +V(\Psi)\label{2.7}
\end{equation}
for the Hamiltonian, this implies
\begin{equation}
\vec{\nabla}_{\Psi}V=0 \notag
\end{equation}
or
\begin{equation}
\frac{\delta V}{\delta \Psi}=0.\label{2.8}
\end{equation}

At a fixed point $\vec{x}_0=[\Psi_0,{\Psi_0}^{\star}]$, the Jacobian matrix of \eqref{2.3} is
\begin{equation}
J(\vec{x}_0)=L \mathscr{H}\label{2.9}
\end{equation}

where
\begin{eqnarray}
\mathscr{H}\equiv \Bigg [\frac{\delta^2 H}{\delta x_i \delta x_j}\Bigg ]_{\vec{x}_0}=
\left(\begin{array}{ccc}
\mathscr{V} & 0\\
0 & I \\
\end{array}\right)\label{2.10}
\end{eqnarray}
from \eqref{2.7}. Here

\begin{equation}
\mathscr{V}=\Bigg [ \frac{\delta^2 V}{\delta \Psi_i \delta \Psi_j}\Bigg ]_{\vec{x}_0}\label{2.11}
\end{equation}
Hence, we have

\begin{eqnarray}
J(\vec{x}_0)=
\left(\begin{array}{ccc}
0 & I\\
-I & 0 \\
\end{array}\right) \left(\begin{array}{ccc}
\mathscr{V}& 0\\
0 & I\\
\end{array}\right)= \left(\begin{array}{ccc}
0 & I\\
-\mathscr{V} & 0 \\
\end{array}\right)\label{2.12}
\end{eqnarray}
whose eigenvalues $\lambda$ satisfy the characteristic equation
\begin{equation}
|\mathscr{V}+\lambda^2 I|=0 \label{2.13}
\end{equation}

Since the matrix $\mathscr{V}$ is symmetric, its eigenvalues are real and the solutions $\lambda$ of \eqref{2.13} are thus either real or purely imaginary. Thus, as claimed earlier, Hopf bifurcations cannot occur in Hamiltonian systems. The introduction of dissipation allows the occurrence of Hopf bifurcation and introduces the various pulsating solitary wave structures which occur in the CGLE.

\subsection{Variational equations}
\subsubsection{Pulsating solitons}

Pulsating solitons  are localized structures with profile that pulsates along the propagation direction  They exist as isolated structures in space and they repeat periodically along $t$ direction.  They have been investigated before in the context of nonlinear optics by \cite{Akhmediev:1,Crespo:1,Mancas:4,Akhmediev:4}. This paper is an extention of our previous work \cite{Mancas:4}  where we analyzed in detail the plain pulsation soliton using a variational mechanism developed by Kaup and Malomed \cite{Kaup:1,Kaup:2,Kaup:3,Kaup:4}. 
 For plain pulsating solitons, the speed is always zero and we take
\begin{equation}\label{4.1}
\phi_1(t)=\phi_2(t)=0.
\end{equation}
%However, we need \underline{not}, in general invoke ~(\ref{3.8}), since the solution of ~(\ref{2.1}) must be complex. T
Therefore, the trial functions ~(\ref{3.2})  become
\begin{align}\label{4.2}
A(x,t)&=A_1(t)e^{-\sigma(t)^2 x^2}e^{i\alpha(t)}\\
r(x,t)&=e^{-\sigma(t)^2x^2}\notag
\end{align}
Substituting into ~(\ref{3.9}), and by choosing $m=1$, the simplified effective  Lagrangian becomes
\begin{eqnarray}\label{4.4}
L_{EFF} & = & \frac{\sqrt{\pi}}{6 \sigma(t)^2}\Bigg[6 A_1(t)^3 \sigma(t) \big(b_3 \cos \alpha(t)+c_3 \sin \alpha(t) \big) \nonumber \\
 &&  +\sqrt 2 \Bigg( 2 \sqrt 3  A_1(t)^5\sigma(t) \big( b_5 \cos \alpha(t)+c_5 \sin \alpha(t) \big)\nonumber \\
 &&  + 6 \dot{A_1}(t) \sigma(t) \cos \alpha(t)- 6 A_1(t) \sigma(t) \sin \alpha(t)\big(c_1\sigma(t)^2+\dot{\alpha}(t)\big)\nonumber \\
 &&  - 3 A_1(t) \cos \alpha(t) \big(\dot{\sigma}(t)+2 \epsilon \sigma(t)-2 b_1 \sigma(t)^3 \big)\Bigg) \Bigg]
\end{eqnarray}
We are left with three ansatz parameters $A_1(t)$,  $\sigma(t)$ and $\alpha(t)$ in $L_{EFF}$ for which we will write the EL variational equations
$$\frac{\partial L_{EFF}}{\partial \star(t)}-\frac{\mathrm{d}}{\mathrm{dt}}\Big(\frac{\partial L_{EFF}}{\partial \dot{\star}(t)}\Big)=0,$$ where $\star$ refers to $A_1$, $\sigma$, or $\alpha$. Solving for $\dot{\star}(t)$ it yields to  a 3D dynamical system
\begin{align}			 
\dot{A_1}(t)&=f_1 \lbrack A_1(t),\sigma(t),\alpha(t) \rbrack \notag\\
\dot{\sigma}(t)&=f_2  \lbrack A_1(t),\sigma(t),\alpha(t) \rbrack \notag\\
\dot{\alpha}(t)&=f_3  \lbrack A_1(t),\sigma(t),\alpha(t) \rbrack, \label{4.5}
\end{align}
where $f_i$, $i=1,2,3$ are complicated nonlinear functions of the arguments, see Fig. \ref{f123} in Appendix.

\subsubsection{Snake solitons}

For this class of solutions, we require the position $\phi_1(t)$ (and phase) to vary \cite{Akhmediev:1}, and we choose choose $m=1$ and $\sigma(t)=\frac{2}{\phi(t)}$, for the 3D dynamical systems. Thus, ~(\ref{3.2})  become
\begin{align}\label{5.2}
A(x,t)&=A_1(t)e^{-\frac{4}{\phi(t)^2}[x-\phi(t)]^2}e^{i\alpha(t)}\\
r(x,t)&=e^{-\frac{4}{\phi(t)^2}x^2}\notag
\end{align}
Substituting into ~(\ref{3.9}), the simplified effective Lagrangian becomes
\begin{eqnarray}\label{5.4}
 L_{EFF} & = & \frac{\sqrt{\pi}}{12 e ^{\frac{10}{3}}\phi(t)}\Bigg[6 e^\frac13 A_1(t)^3 \phi(t)^2 \big(b_3 \cos \alpha(t)+c_3 \sin \alpha(t) \big) \nonumber \\
 &&  + 2 \sqrt 6  A_1(t)^5\phi(t)^2 \big( b_5 \cos \alpha(t)+c_5 \sin \alpha(t) \big)\nonumber \\
 &&  -3 \sqrt 2e^{-\frac43}\bigg(-2 A_1(t)\sin \alpha(t)\big(-12 c_1+\phi(t)^2\dot{\alpha(t)}\big)\nonumber \\
 &&  +  \cos \alpha(t) \big(-2 \phi^2(t)\dot{\alpha}(t)+A_1(t)(24 b_1+2 \epsilon\phi^2(t)+3\phi(t)\dot{\phi}(t)) \big)\bigg) \Bigg]
\end{eqnarray}
As in the previous case, we are left with three parameters $A_1(t)$,  $\phi(t)$ and $\alpha(t)$ in $L_{EFF}$. 
Varying them we obtain
$$\frac{\partial L_{EFF}}{\partial \star(t)}-\frac{\mathrm{d}}{\mathrm{dt}}\Big(\frac{\partial L_{EFF}}{\partial \dot{\star}(t)}\Big)=0,$$ where $\star$ refers to $A_1$, $\phi$, or $\alpha$. Solving for $\dot{\star}(t)$ the 3D system is, 
\begin{align}			 
\dot{A_1}(t)&=f_4 \lbrack A_1(t),\phi(t),\alpha(t) \rbrack \notag\\
\dot{\phi}(t)&=f_5  \lbrack A_1(t),\phi(t),\alpha(t) \rbrack \notag\\
\dot{\alpha}(t)&=f_6  \lbrack A_1(t),\phi(t),\alpha(t) \rbrack, \label{5.5}
\end{align}
where $f_i$, $i=4,5,6$ are complicated nonlinear functions of the arguments, see Fig. \ref{f456} in Appendix.

\subsection{Numerical simulations}
Using the variational formulation, and by varying the parameters of the ansatz ~(\ref{4.2}), ~(\ref{5.2}), we obtain two systems of highly nonlinear ODEs, \eqref{4.5}, \eqref{5.5} which can be written in compact form as
\begin{eqnarray}\label{eq5}
&\frac{d {\vec u}}{d t}={\vec F}\left({\vec u},{\vec p}\right),\,  t\in
[0,T]; \,\, {\vec u}(0)=\vec u_0, &
\end{eqnarray}
where ${\vec u}=(A_1,\sigma,\alpha), \, {\vec F}=(f_{1},f_{2},f_{3})$ for pulses,  ${\vec u}=(A_1,\phi,\alpha), \, {\vec F}=(f_{4},f_{5},f_{6})$ for snakes, and ${\vec p}=(b_1, b_3, b_5, c_1, c_3, c_5, \epsilon)$ is a constant vector that depends on the system parameters listed in the Table \ref{tab:1}.

\begin{table}
\caption{Parameters of the CGLE}
\label{tab:1}       % Give a unique label
\begin{tabular}{llllllll}
\hline Solitons & $\epsilon$ & $b_1$ & $c_1$ & $b_3$ & $c_3$ & $b_5$ & $c_5$ \\
\hline
pulses & -0.100 & 0.080 & 0.500 & -0.660 & 1.000 & 0.100 & -0.100  \\
\hline
snakes & -0.100 & 0.080 & 0.500 & -0.835 & 1.000 & 0.110 & -0.080  \\
\hline
\end{tabular}
\label{tabu}
\end{table}
For our purposes the ODEs have been  solved numerically for different parameters $\vec p$ using RK4 method with constant initial conditions:  $A_1(0) = A_0$, $\sigma(0) =\sigma_0$, $\alpha(0)=\alpha_0$ for pulses and  $A_1(0) = A_0$, $\phi(0) =\phi_0$, $\alpha(0)=\alpha_0$ for snakes. The initial conditions ${\vec u_0}$ used were the fixed points of the EL equations that we found numerically by solving the algebraic system ${\vec F}\left({\vec u},{\vec p}\right)={\vec 0}$ of transcendental equations.

\subsection{Hopf bifurcations}

We derive the conditions for the temporal Hopf bifurcations of the fixed points. The conditions for supercritical temporal Hopf bifurcations, leading to stable periodic orbits of $A_1(t)$, $\sigma(t)$ or $\phi(t)$, and $\alpha(t)$ are evaluated using the method of multiple scales from Sect. ~\ref{sec:4}. These are the conditions or parameter regimes where they exhibit stable periodic oscillations, and hence stable pulsating solitons and snakes will exist within our variational formulation. Note that periodic oscillations of $A_1(t)$, $\sigma(t)$ or $\phi(t)$, and $\alpha(t)$, correspond to a spatiotemporal pulsating soliton structure of the $|A(x,t)|$.

For a typical fixed point ${\vec u_0}$ , the characteristic polynomial of the Jacobian matrix  of   ~(\ref{4.5}), ~(\ref{5.5}) is 
\begin{equation} \label{4.6}
\lambda^3+\delta_1\lambda^2+\delta_2\lambda +\delta_3=0
\end{equation}
where
$\delta_i$ with $i=1,2,3$ depend on both  the system parameters ${\vec p}$ and the fixed points $\vec u_0$ . To be a stable fixed point within the linearized analysis, all the eigenvalues must have negative real parts. Using the Routh-Hurwitz criterion, the necessary and sufficient conditions for ~(\ref{4.6}) to have $Re(\lambda_{1,2,3})<0$ are:
\begin{equation}\label{4.7}
\delta_1>0,\quad\delta_3>0,\quad\delta_1\delta_2-\delta_3>0.
\end{equation}

On the contrary, one may have the onset of instability of the plane wave solution occurring in one of two ways. In the first case, one eigenvalue of the Jacobian becomes non-hyperbolic by going through zero when 
\begin{equation} \label{4.8}
\delta_3=0.
\end{equation}
Equation ~(\ref{4.8}) is thus the condition for the onset of ``static" instability of the plane wave. Whether this bifurcation is a pitchfork or transcritical one, and its subcritical or supercritical nature, may be readily determined by deriving an appropriate canonical system in the vicinity of ~(\ref{4.8}) using any variety of normal forms or perturbation methods.

The second  dynamic instability  is when a pair of eigenvalues of the Jacobian become purely imaginary. The consequent Hopf bifurcation at
\begin{equation}\label{4.9}
\delta_1\delta_2-\delta_3=0
\end{equation}
leads to the onset of periodic solutions of the dynamical systems ~(\ref{4.5}), ~(\ref{5.5}) (dynamic instability or ``flutter''). 

\subsection{Effects of system parameters on the shape of the solitons}

Within the regimes of stable periodic solutions, we  investigate:\\
a) the effects of the system\rq{}s parameters on the shape and structure of the solitons, and\\
b) the period doubling sequences as the above system parameters are varied. 

To study the effects of system parameters on the shape and the stability of the solitons, we integrate ~(\ref{4.5}),~(\ref{5.5}) numerically for different sets of the various system parameters  within the regime of stable periodic solutions. The resulting periodic time series for $A_1(t)$, $\sigma(t)$ or $\phi(t)$ and $\alpha(t)$  are substituted in ~(\ref{4.2}), and ~(\ref{5.2}) whose spatiotemporal structure is  plotted. As the various system parameters within the stable regime are varied, the effects of the soliton amplitude, width, and phase are studied.

\section{Stability Analysis of Periodic Orbits}\label{sec:4}
We develop a generalized 3D multiple scales method to construct analytical approximations for the periodic orbits \cite{Nody1} arising through Hopf bifurcation of the fixed point of the EL equations ~(\ref{4.5}), ~(\ref{5.5}). For both systems the limit cycle is determined by expanding the amplitude $A_1(t)$, width $\sigma(t)$ or speed $\phi(t)$, and phase $\alpha(t)$, using progressively slower spatial scales.

In the standard way, we write the various or multiple scales as $z=Z_0$, $Z_1=\delta Z_0$, $Z_2=\delta^2 Z_0$, $\cdots$, where $\delta$ is the usual multiple scales expansion parameter. We shall expand in powers of $\delta$, to separate the various scales, and then set $\delta =1$ at the end. We chose the cubic gain parameter $b_3$, as the control or bifurcation parameter. The expansion takes the form (throughout the section for snakes one shall use expansion in  $\phi$ instead of $\sigma$)
\begin{align}
A_1&=A_{11}(Z_0,Z_1,Z_2)+\delta A_{12}(Z_0,Z_1,Z_2)+ \delta^2 A_{13}(Z_0,Z_1,Z_2)\cdots \label{3.2.1},\\
\sigma&=\sigma_1(Z_0,Z_1,Z_2)+\delta \sigma_2(Z_0,Z_1,Z_2)+\delta^2 \sigma_3(Z_0,Z_1,Z_2)\cdots\label{3.2.2},\\
\alpha&=\alpha_1(Z_0,Z_1,Z_2)+\delta \alpha_2(Z_0,Z_1,Z_2)+\delta^2 \alpha_3(Z_0,Z_1,Z_2)\cdots.\label{3.2.3}
\end{align}
Using the chain rule, the spatial derivative becomes
\begin{equation}\label{3.2.4}
\frac{\mathrm{d}}{\mathrm{d}Z}=D_0+\delta D_1+\delta^2 D_2+\cdots,
\end{equation}
where $D_n=\frac{\partial}{\partial Z_n}$. The delay parameter $b_3$ is ordered as 
\begin{equation}\label{3.2.5}
b_3=b_{30}+\delta^2 b_{32},
\end{equation}
where $b_{30}$ is the critical value such that ~(\ref{4.7}) is not satisfied, (i.e. $b_{30}$ is a solution of
~(\ref{4.9}). This is standard for this method \cite {Dodd}, as it allows the influence from the nonlinear terms and the control parameter to occur at the same order.

Using ~(\ref{3.2.1})-~(\ref{3.2.5}) and equating like powers of $\delta$ yields equations at $\mathrm{O}(\delta^i)$ of the form:
\begin{eqnarray}
\frac{d}{d Z_0} \vec{x}_i+\left(\begin{array}{ccc}
f_{1v} & f_{2v} & f_{3v}\\
f_{1w} & f_{2w} & f_{3w}\\
f_{1z} & f_{2z} & f_{3z}\\
\end{array}\right)  \vec{x}_i=\vec{S}_{i,j} \label{6.5}
\end{eqnarray} where, $i=1,2,3$, represents the order, $j=1,2,3$ represents the equations' number, and $\vec S_{i,j}$ is the source or inhomogeneous term for the $j^\mathrm{th}$ equation at $\mathrm{O}(\delta^i)$,
\begin{eqnarray}
\vec{x}_i=\left(\begin{array}{ccc}
A_{1i}(Z_0,Z_1,Z_2)\\
\sigma_i(Z_0,Z_1,Z_2)\\
\alpha_i(Z_0,Z_1,Z_2)\\
\end{array}\right).\notag
\end{eqnarray}
Here,
\begin{eqnarray}
\left(\begin{array}{ccc}
f_{1v} & f_{2v} & f_{3v}\\
f_{1w} & f_{2w} & f_{3w}\\
f_{1z} & f_{2z} & f_{3z}\\
\end{array}\right) = J\Big[{\frac{\partial f_1,\partial f_2,\partial f_3}{\partial A_1,\partial \sigma,\partial \alpha}}\Big]\label{6.6}
\end{eqnarray}
where $J$ is the Jacobian matrix of ~(\ref{4.5}), ~(\ref{5.5}),  evaluated numerically at the fixed points $\vec u_0$. 
For all orders, the structure of the equations is the same, only the source terms $\vec S_{i,j}$ are different, and they are represented below order by order.
$$\mathrm{O}(\delta^1):$$
\begin{align}
S_{1,j}&=0 \label{A.11}
\end{align}
$$\mathrm{O}(\delta^2):$$
\begin{align}
S_{2,1}&=\frac12(f_{1vv}A_{11}^2+f_{1ww}\sigma_{1}^2+f_{1zz}\alpha_{1}^2)\label{A.21}\\
&+f_{1vw}A_{11}\sigma_1+f_{1vz}A_{11}\alpha_1+f_{1wz}\sigma_{1}\alpha_1-2 D_1A_{11}\notag\\
S_{2,2}&=\frac12(f_{2vv}A_{11}^2+f_{2ww}\sigma_{1}^2+f_{2zz}\alpha_{1}^2)\label{A.22}\\
&+f_{2vw}A_{11}\sigma_1+f_{2vz}A_{11}\alpha_1+f_{2wz}\sigma_{1}\alpha_1-2 D_1\sigma_1\notag\\
S_{2,3}&=\frac12(f_{3vv}A_{11}^2+f_{3ww}\sigma_{1}^2+f_{3zz}\alpha_{1}^2)\label{A.23}\\
&+f_{1vw}A_{31}\sigma_1+f_{3vz}A_{11}\alpha_1+f_{3wz}\sigma_{1}\alpha_1-2 D_1\alpha_1\notag
\end{align}
$$\mathrm{O}(\delta^3):$$
\begin{align}
S_{3,1}&=\frac16(f_{1vvv}A_{11}^3+f_{1www}\sigma_{1}^3+f_{1zzz}\alpha_{1}^3)\label{A.31}\\
&+\frac12(f_{1vvw}A_{11}^2\sigma_1+f_{1vvz}A_{11}^2\alpha_{1}+f_{1wwz}\sigma_{1}^2\alpha_1+f_{1vzz}A_{11}\alpha_1^2+f_{1wzz}\sigma_1\alpha_1^2+f_{1vww}A_{11}\sigma_1^2)\notag\\
&+g_{1v}A_{11}+g_{1w}\sigma_1+g_{1z}\alpha_1+f_{1vv}A_{11}A_{12}+f_{1ww}\sigma_1\sigma_2+f_{1zz}\alpha_1\alpha_2\notag\\
&+f_{1vz}(A_{11}\alpha_2+A_{12}\alpha_1)+f_{1vw}(A_{11}\sigma_2+A_{12}\sigma_1)+f_{1wz}(\sigma_1\alpha_2+\sigma_2\alpha_1)\notag\\
&+f_{1wz}A_{11}\sigma_1\alpha_1-D_2A_{11}-D_1A_{12}\notag
\end{align}
\begin{align}
S_{3,2}&=\frac16(f_{2vvv}A_{11}^3+f_{2www}\sigma_{1}^3+f_{2zzz}\alpha_{1}^3)\label{A.32}\\
&+\frac12(f_{2vvw}A_{11}^2\sigma_1+f_{2vvz}A_{11}^2\alpha_{1}+f_{2wwz}\sigma_{1}^2\alpha_1+f_{2vzz}A_{11}\alpha_1^2+f_{2wzz}\sigma_1\alpha_1^2+f_{2vww}A_{11}\sigma_1^2)\notag\\
&+g_{2v}A_{11}+g_{2w}\sigma_1+g_{2z}\alpha_1+f_{2vv}A_{11}A_{12}+f_{2ww}\sigma_1\sigma_2+f_{2zz}\alpha_1\alpha_2\notag\\
&+f_{2vz}(A_{11}\alpha_2+A_{12}\alpha_1)+f_{2vw}(A_{11}\sigma_2+A_{12}\sigma_1)+f_{2wz}(\sigma_1\alpha_2+\sigma_2\alpha_1)\notag\\
&+f_{2wz}A_{11}\sigma_1\alpha_1-D_2\sigma_1-D_1\sigma_2\notag
\end{align}
\begin{align}
S_{3,3}&=\frac16(f_{3vvv}A_{11}^3+f_{3www}\sigma_{1}^3+f_{3zzz}\alpha_{1}^3)\label{A.33}\\
&+\frac12(f_{3vvw}A_{11}^2\sigma_1+f_{3vvz}A_{11}^2\alpha_{1}+f_{3wwz}\sigma_{1}^2\alpha_1+f_{3vzz}A_{11}\alpha_1^2+f_{3wzz}\sigma_1\alpha_1^2+f_{3vww}A_{11}\sigma_1^2)\notag\\
&+g_{3v}A_{11}+g_{1w}\sigma_1+g_{3z}\alpha_1+f_{3vv}A_{11}A_{12}+f_{3ww}\sigma_1\sigma_2+f_{3zz}\alpha_1\alpha_2\notag\\
&+f_{3vz}(A_{11}\alpha_2+A_{12}\alpha_1)+f_{3vw}(A_{11}\sigma_2+A_{12}\sigma_1)+f_{3wz}(\sigma_1\alpha_2+\sigma_2\alpha_1)\notag\\
&+f_{3wz}A_{11}\sigma_1\alpha_1-D_2\alpha_1-D_1\alpha_2\notag
\end{align}
Here, the $g_i$ functions are obtained by using ~(\ref{3.2.5}) in $f_i$ as this variation will introduce additional terms of higher order. i.e. $f_i \rightarrow f_i+ \delta^2 g_i$. So the new $f_i$ will contain $b_{30}$ terms and represents the fact that we are situated on the Hopf bifurcation curve, while $g_i$'s contain $b_{32}$ terms, and shows how far we are from the curve.

Now we will proceed to solve ~(\ref{6.5}) order by order. Since the sources for the first order system are identically zero, we may assume the first order solution of ~(\ref{6.5}) to be 
\begin{eqnarray}
\vec{x}_1=\left(\begin{array}{ccc}
\beta_1\\
\gamma_1\\
\eta_1\\
\end{array}\right)e^{-i \omega_0 Z_0}+ c.c. ,\label{6.7}
\end{eqnarray}
and substituting back this solution into ~(\ref{6.5}), we obtain
the eigenvalue  $\omega_0$, with corresponding eigenvector $\vec{x}_1$.
By looking at the characteristic polynomial of the Jacobian matrix of ~(\ref{4.6}) we obtain  
\begin{equation}
\delta_2=\omega_0^2=-f_{1w}f_{2v}+f_{1v}f_{2w}-f_{1z}f_{3v}+f_{1v}f_{3z}+f_{2w}f_{3z}.\label{cond1}
\end{equation}
Hence, the first order solution of ~(\ref{6.5}), $\vec{x}_1$ can be written as 
\begin{align}
A_{11}&=(a+i b)\theta(Z_1,Z_2)e^{i \omega_0 Z_0}+(a-i b)\bar{\theta}(Z_1,Z_2)e^{-i \omega_0 Z_0}\label{6.8}\\
\sigma_1&=(c+i d)\theta(Z_1,Z_2)e^{i \omega_0 Z_0}+(c-i d)\bar{\theta}(Z_1,Z_2)e^{-i \omega_0 Z_0}\label{6.9}\\
\alpha_1&=\theta(Z_1,Z_2)e^{i \omega_0 Z_0}+\bar{\theta}(Z_1,Z_2)e^{-i \omega_0 Z_0},\label{6.44}
\end{align}
where $\eta_1$ is taken to be $1$, $\beta_1 \equiv a+i b$, and $\gamma_1 \equiv c+ i d$. Now, since the first order solutions ~(\ref{6.8}), ~(\ref{6.9}), and ~(\ref{6.44}) are known, the second order sources $S_{2,j}$ are evaluated via ~(\ref{A.21})-~(\ref{A.23}) which take the form 

\begin{eqnarray}
\vec{S}_{2,j}=\left(\begin{array}{ccc}
S_{2,1}^{(0)}\\
S_{2,2}^{(0)}\\
S_{2,3}^{(0)}\\
\end{array}\right)+ \left(\begin{array}{ccc}
S_{2,1}^{(1)}\\
S_{2,2}^{(1)}\\
S_{2,3}^{(1)}\\
\end{array}\right)e^{i \omega_0 Z_0}+ \left(\begin{array}{ccc}
S_{2,1}^{(2)}\\
S_{2,2}^{(2)}\\
S_{2,3}^{(2)}\\
\end{array}\right)e^{2i \omega_0 Z_0}+c.c. ,\label{6.10}
\end{eqnarray}
Setting the coefficients of the secular first harmonic or $e^{i \omega_0 Z_0}$ terms (which are the solutions of the homogeneous equation) to zero, i.e. $\vec{S}_{2,j}^{(1)}=\vec{0}$ yields
\begin{align}
D_1\theta &=\frac{\partial \theta}{\partial Z_1}=0\label{6.11}\\
D_1\bar{\theta} &=\frac{\partial \bar{\theta}}{\partial Z_1}=0.\notag
\end{align}
Using ~(\ref{6.11}), ~(\ref{6.8})-~(\ref{6.10}), and the second order sources ~(\ref{6.10}), and by assuming a second order particular solution of ~(\ref{6.5}) of the type
\begin{eqnarray}
\vec{x}_2=\left(\begin{array}{ccc}
A_{12}^{(0)}\\
\sigma_2^{(0)}\\
\alpha_2^{(0)}\\
\end{array}\right)+ \left(\begin{array}{ccc}
A_{12}^{(2)}\\
\sigma_2^{(2)}\\
\alpha_2^{(2)}\\
\end{array}\right)e^{2i \omega_0 Z_0}+c.c. ,\label{6.12}
\end{eqnarray}
we solve the system ~(\ref{6.5}) for the unknowns $A_{12}^{(0)}$, $\sigma_2^{(0)}$, and $\alpha_2^{(0)}$, by looking at the homogeneous system, and for the unknowns $A_{12}^{(2)}$, $\sigma_2^{(2)}$, and $\alpha_2^{(2)}$, by looking at the inhomogeneous system ~(\ref{6.5}). Using the full second order solution $\vec{x}_2$, which includes the DC terms and the second harmonic terms, and the previously found first order solution $\vec{x}_1$, we can find the third order sources via ~(\ref{A.31})-~(\ref{A.33}). By writing the third order sources as 

\begin{eqnarray}
\vec{S}_{3,j}=\left(\begin{array}{ccc}
S_{3,1}^{(0)}\\
S_{3,2}^{(0)}\\
S_{3,3}^{(0)}\\
\end{array}\right)+ \left(\begin{array}{ccc}
S_{3,1}^{(1)}\\
S_{3,2}^{(1)}\\
S_{3,3}^{(1)}\\
\end{array}\right)e^{i \omega_0 Z_0}+\\ \left(\begin{array}{ccc}
S_{3,1}^{(2)}\\
S_{3,2}^{(2)}\\
S_{3,3}^{(2)}\\
\end{array}\right)e^{2i \omega_0 Z_0}+ \left(\begin{array}{ccc}
S_{3,1}^{(3)}\\
S_{3,2}^{(3)}\\
S_{3,3}^{(3)}\\
\end{array}\right)e^{3i \omega_0 Z_0}+ c.c. ,\label{6.13}
\end{eqnarray}
we can find the coefficient of the secular terms  $e^{i \omega_0 Z_0}$ terms, i.e. $\vec{S}_{3,j}^{(1)}$.
Now, the evolution equation can be found by solving ~(\ref{6.14}).
\begin{eqnarray}
\left (\begin{array}{ccc}
f_{1v}+i \omega_0 & f_{2v} & f_{3v}\\
f_{1w} & f_{2w}+i \omega_0 & f_{3w}\\
f_{1z} & f_{2z} & f_{3z}+i \omega_0 \\
\end{array}\right)  \vec{x}_3=\vec{S}_{3,j}^{(1)} \label{6.14}
\end{eqnarray}
This system can be written in the compact form 
\begin{equation}
(\textbf{A}-\lambda \textbf{I})\vec{x}_3=\vec{S}_{3,j}^{(1)},\label{6.50}
\end{equation} where $\lambda=\pm i \omega_0$ are the eigenvalues of $\textbf{A}.$
By the Fredholm alternative, ~(\ref{6.50}) has solution iff $\vec{S}_{3,j}^{(1)}\in Range(\textbf{A}-\lambda \textbf{I})$. The final evolution equation for the coefficients \cite{Nayfeh} in the linear solutions of ~(\ref{6.5}) is obtained from
\begin{eqnarray}
\left |\begin{array}{ccc}
S_{3,1}^{(1)} & f_{2v} & f_{3v}\\
S_{3,2}^{(1)} & f_{2w}+i \omega_0 & f_{3w}\\
S_{3,3}^{(1)} & f_{2z} & f_{3z}+i \omega_0 \\
\end{array}\right|=0 \label{6.15}
\end{eqnarray}
From ~(\ref{6.15}), we have the evolution equation on the slow second order $Z_2$ scale
\begin{equation}\label{6.16}
\frac{\partial\theta}{\partial{Z_2}}=S_1\theta^2\bar{\theta}+S_2\theta.
\end{equation}
Writing $\theta=\frac12 A e^{i\zeta}$ and separating ~(\ref{6.15}) into real and imaginary parts, yields
\begin{equation}\label{6.17}
\frac{\partial A}{\partial{Z_2}}=\frac{S_{1r}A^3}{4}+S_{2r}A,
\end{equation}
where $S_{1r}$ and $S_{2r}$ represent the real parts of $S_1$ and $S_2$ respectively.
In the usual way, the fixed points of ~(\ref{6.15}), $(A_1,A_{2,3})$ where
\begin{align}
A_1&=0,\notag\\
A_{2,3}&=\pm 2\sqrt{-\frac{S_{2r}}{S_{1r}}} \label{6.18}
\end{align}
give the amplitude of the solution $\theta=\frac12Ae^{i\zeta}$, with $A_{2,3}$ corresponding to the bifurcation periodic orbits. Clearly $A_{2,3}$ are real fixed points whenever 
\begin{equation}\label{6.19}
\frac{S_{2r}}{S_{1r}}<0,
\end{equation}
and the Jacobian of the right hand side of ~(\ref{6.19}) evaluated at $A_{2,3}$ is $J|_{A_{2,3}}=-2S_{2r}$, where $J(A)=\frac{\partial{(\frac{S_{1r}A^3}{4}+S_{2r}A})}{\partial A}$. Clearly, a necessary condition for stability is to have $S_{2r}>0$, and for instability $S_{2r}<0$. Thus, the system undergoes:
\begin{itemize}
\item[a)] supercritical Hopf bifurcations when 
\begin{equation}\label{super}
S_{2r}>0, \qquad S_{1r}<0,
\end{equation}
\item[b)]	subcritical Hopf bifurcations when
\begin{equation}
S_{2r}<0, \qquad S_{1r}>0.
\end{equation}
\end{itemize}

We  use ~(\ref{super}) next to identify regimes of supercritical bifurcations where the solutions of the EL equations ~(\ref{4.5}) or ~(\ref{5.5}) for pulsating or snake solitons will result in oscillations of $A_1(t)$, $\sigma(t)$ or $\phi(t)$ and $\alpha(t)$ that when substituted into ansatz will lead to pulsating and snake solitons.

\section{Results for the Plane Pulsating Soliton}\label{sec:5}

An example of a plain pulsating solitonobtained using the RK4 method on \eqref{eq5} in Fortran is shown in Fig. \ref{Set1} using the ansatz ~(\ref{4.2}). It has a different shape at each time $t$, since it evolves, but it recovers its exact initial shape after a period.

To derive the conditions for occurrence of stable periodic orbits of $A_1(t)$, $\sigma(t)$, and $\alpha(t)$, first we fix a set of  system parameters $b_1=0.08$, $b_5= 0.1$, $c_1=0.5$, $c_3=1$, $c_5=-0.1$, see line 1 of Table \ref{tab:1}.  Then, we find numerically the fixed points of  ~\ref{4.5}. By the Ruth-Hurwitz conditions, the Hopf curve is defined as $\delta_1 \delta_2-\delta_3=0.$ This condition, along with the equations of the fixed points leads to onset of periodic solutions of ~(\ref{4.5}).

On the Hopf bifurcation curve we find that $b_3=-0.216825$, and $\epsilon=-0.345481$, while the fixed points are 
$ \vec{u}_0=\left(A_1(0),\sigma(0),\alpha(0)\right)=\left(0.954712, 0.917093,-0.181274 \right)$
Using these values of $b_3$ and $\epsilon$,  Hopf bifurcations occur in this system leading to periodic orbits \cite{Nody1}.

Next, we  plot the time series of the periodic orbit for the amplitude $A_1(t)$, and as expected, we noticed that the amplitude was very small, since it is proportional to the square root of the distance from the Hopf curve. 

To construct pulsating solitons with amplitudes large enough, one has move away from the Hopf curve while staying inside of the parameters ranges for the existence of the pulsating soliton. First, we varied $\epsilon$ slowly away from the Hopf curve. Repeating the above procedure to construct a pulsating soliton, we noticed that it still had very small amplitudes $A_1(t)$, of magnitude only of $10^{-4}$. Hence, we decided to also vary the bifurcation  parameter $b_3$ instead. We found that the domain of existence for the pulsating soliton as  $b_3$ varied was $[-0.2531943,-0.1424]$, passing through the Hopf curve value of $b_{3Hopf}=-0.216825$. Within this range, we studied the effects on the shape and the stability, as well as the various bifurcations that lead potentially to period doubling and quadrupling. For the largest value of $b_3$, i.e. $b_3=-0.1424$, we numerically integrate using RK4 method on \eqref{4.5}, and we plot the periodic orbit, which is shown in Fig. \ref{Figure22}.

The resulting periodic time series for $A_1(t)$, $\sigma(t)$, and $\alpha(t)$ from Fig. \ref{Figure22} are substituted in \eqref{4.2} whose spatiotemporal structure  is  plotted in Fig. \ref{Figure22}. We continued with the rest of the parameters $c_1$, $c_3$, $c_5$, $b_1$, $b_5$ within the stable regime and  the effects on  amplitude, width, position, phase speed (and, less importantly, phase) were analyzed. We also showed the orbit and the plane pulsating soliton for the smallest value of $b_3=-0.2531943$ in Fig. \ref{Figure25}.

Next, we consider the detailed effects of varying the parameter $b_3$. For the chosen values of the system parameters of $b_1=0.08$, $b_5= 0.1$, $c_1=0.5$, $c_3=1$, $c_5=-0.1$, and $\epsilon=-0.345481$, the Hopf bifurcation occurs at $b_{3Hopf}=-0.216825$

First, let us consider values of $b_3>b_{3Hopf}$. There is a stable and robust periodic orbit to this side which becomes larger and deforms as $b_3$ is increased up to $-0.1424$. A representative periodic orbit is in Fig. \ref{Figure22}.

Next, moving to values smaller than $b_{3Hopf}$, we see a clean, periodic orbit which slowly grows in size as $b_3$ is made more negative. The periodic orbit, time series, and solitary waves are qualitatively similar to those for $b_3>b_{3Hopf}$.

However, more interesting dynamics is seen as $b_3$ is decreased further. The periodic orbit goes unstable via a very rapid, complete cascade of period-doubling bifurcations between $b_3=-0.25$, and $b_3=-0.2516$. In Fig. \ref{Figure27} we show the period doubled orbit for $b_3=-0.2516$. The orbit at $b_3=-0.2531943$ after many more period doublings is shown in Fig. \ref{Figure25}. The corresponding solitary wave solution is shown in Fig. \ref{Figure25}. Note also that one may track the complete cascade of period doublings using software such AUTO or DERPER, or using the schemes of Holodniok and Kubicek \cite{Holodniok}.

Next, we explain effect of all the various parameters in the CGLE \eqref{2.1} on the shape (amplitude, width, period) and stability of the pulsating solitary wave. 
In considering the parameter effects on the solitary wave shape and period, note that the wave is a spatially coherent structure (or a ``collective coordinate" given by the trial function) whose parameters oscillate in time. Hence, the temporal period of the pulsating soliton is the same as the period $T$ of the oscillations of $A_1(t)$, $\sigma(t)$, and $\alpha(t)$ on their limit cycle. As for the peak amplitude and peak width of the pulsating wave, these are determined by the peak amplitude $A_{1p}$ of $A_1(t)$, and the reciprocal of the peak amplitude $\sigma_p$ of $\sigma(t)$ respectively, i.e. at any time $t$ when the amplitude is maximum, the width will be minimum, and vice versa. 

Keeping the above in mind, we varied the parameters of the CGLE in turn and observed the resulting effects on $A_{1p}$ (the peak amplitude), $\sigma_p$ (the inverse width), and $T$ (the temporal period) of the pulsating soliton:\\
(i) For \textit{increased} $b_1$, the values of $A_{1p}$, $\sigma_p$, and $T$ all \textit{increase}.\\
(ii) \textit{Increasing} $b_5$ \textit{augments} all of $A_{1p}$, $\sigma_p$, and $T$.\\
(iii) \textit{Raising} $c_1$ \textit{increases } $A_{1p}$, $\sigma_p$, and $T$.\\
(iv) \textit{Incrementing} $c_3$ \textit{decreases} all of $A_{1p}$, $\sigma_p$, and $T$.\\
(v) \textit{Augmenting} $c_5$ causes a \textit{decrease} in $A_{1p}$, $\sigma_p$, and $T$.\\
(vi) \textit{Raising} $\epsilon$ causes  $A_{1p}$, $\sigma_p$, and $T$ to \textit{fall}. 
These results can be seen in Figs.\ref{plane10},\ref{plane12}.
The results in cases (a),(c),(e) of Figs. \ref{plane10}, \ref{plane12} are to be compared with the plane pulsating soliton obtained by numerical simulations from Fig. \ref{Set1}. The results in cases (b),(d),(f) of Figs. \ref{plane10}, \ref{plane12} are to be compared with the plane pulsating soliton obtained by variational approximation from Fig. \ref{Figure25}. The above analysis constitutes our detailed predictions of the various parameters in the CGLE on the amplitude, inverse width, and temporal width of the pulsating solitons. We have verified that each set of predictions (a)-(f) above agree when the corresponding parameter is varied in the solitary wave simulation shown in Fig. \ref{Set1}. Note also that $A_1(t)$ and $\sigma(t)$ are always in phase, so that $A_{1p}$ and $\sigma_p$ occur simultaneously. Thus, the pulsating solitons are tallest where they have least width. This is also  consistent with the simulations of \cite{Artigas,Akhmediev:1,Crespo:1}.

\section{Results for the Snake Solitons}\label{sec:6}

An example of a snake soliton is shown in Fig. \ref{Set2} using the trial functions ~(\ref{5.2}). The soliton  would now ``snake'' or wiggle as its position varies periodically. Note that the amplitude of the field $|A(x,t)|$ varies periodically as $A_1(t)$ varies, but there would be additional amplitude modulation due to the periodic variation of $\phi(t)$.

To derive the conditions for occurrence of stable periodic orbits of $A_1(t)$, $\phi(t)$, and $\alpha(t)$, we proceeded as follows.

First, we fixed the set of parameters $b_1=0.08$, $b_5= 0.11$, $c_1=0.5$, $c_3=1$, $c_5=-0.08$.  Then, we solved numerically the system of transcendental equations ~(\ref{5.5}). On the Hopf bifurcation curve we got that $b_{3Hopf}=-1.89646$, and $\epsilon=-0.297393$, while the fixed points were $ \vec{u}_0=\left(A_1(0),\phi(0),\alpha(0)\right)=\left(0.583236, 1.05969, 0.185515 \right)$.  To construct snake solitons with amplitudes large enough, we had to move again  away  from the Hopf curve.  For the value of $b_3=-0.835$, and $\epsilon=-0.1$ we numerically integrate using RK4 method the three differential equations ~\eqref{5.5}. The resulting periodic time series for $A_1(t)$, $\phi(t)$, and $\alpha(t)$ were inserted in \eqref{5.2} whose spatiotemporal structure is  plotted. Fig. \ref{Set6} shows the periodic time series for amplitude $A_1(t)$, position $\phi(t)$, and phase  $\alpha(t)$ for $b_3=-0.835$. When we increase $b_3$ to $-0.61$
the oscillations are slightly increasing while the period is decreasing, so the snakes wiggles more, see Fig. \ref{Set7}. Furthermore, when we increase $\epsilon$ to $-0.08$ we see an accentuated decrease in period, see Fig. \ref{Set8}.

In considering the parameter effects on snake shape and period, note that the snake is a spatially coherent structure (or a ``collective coordinate" given by the trial function) whose parameters oscillate in time. Hence, the temporal period of the snake is the same as the period $T$ of the oscillations of $A_1(t)$, $\phi(t)$, and $\alpha(t)$ on their limit cycle. As for the peak amplitude and peak position of the snake, these are determined by the peak amplitude $A_{1p}$ of $A_1(t)$, and the peak position $\phi_p$ of $\phi(t)$ respectively. Notice that from ~(\ref{5.2}) we can regard the width and the amplitude of the snake as being inverse proportional with position $\phi(t)$ for the snake i.e., at any time $t$ when the amplitude is minimum, the width will be minimum, so the position is maximum and vice versa. So, maximum deflection from the horizontal position $x=const.$ is obtained when the position of the snake is maximum, and hence the width and amplitude are minimum. This can be clearly seen in Figs. \ref{snake12}, ~\ref{snake13}.

Keeping the above in mind, we varied  the parameters of the CGLE and  we observed the resulting effects on $A_{1p}$ (the peak amplitude), $\phi_p$ (the position), and $T$ (the temporal period) of the snake soliton:\\
(vii) For \textit{increased} $b_1$, the values of $A_{1p}$, $\phi_p$, and $T$ all \textit{increase}.\\
(viii) \textit{Increasing} $b_3$ \textit{augments} all of $A_{1p}$, $\phi_p$, and $T$.\\
(ix) \textit{Increasing} $b_5$ \textit{increases} all of $A_{1p}$, $\phi_p$, and $T$.\\
(x) \textit{Raising} $c_1$ \textit{increases } $A_{1p}$, $\phi_p$, but \textit{decreases} $T$.\\
(xi) \textit{Incrementing} $c_3$ \textit{decreases} all of $A_{1p}$, $\phi_p$, and $T$.\\
(xii) \textit{Augmenting} $c_5$ causes a \textit{decrease} in $A_{1p}$, $\phi_p$, and \textit{increases} $T$.\\
(xiii) \textit{Raising} $\epsilon$ causes  $A_{1p}$, $\sigma_p$ to \textit{rise}, but $T$ to \textit{fall}.

%%%%%%%%%%%%%%%%%%%%%%%%%%%%%%%%%%%%%%%
The  above above analysis constitutes our detailed predictions of the various parameters in the CGLE on the amplitude, position, and temporal width of the snake solitons. We have verified that each set of predictions (g)-(m) above agree when the corresponding parameter is varied in the solitary wave simulation  shown in Fig. \ref{Set2}. Note also that $A_1(t)$ and $\phi(t)$ are always in phase, so that $A_{1p}$ and $\phi_p$ occur simultaneously. Thus, the pulsating solitons are tallest where they have most width. This is completely consistent with our simulation in Fig. \ref{Set2}, as well as those in \cite{Artigas}.

\section{Conclusions}\label{sec:8}

In this article we have developed a comprehensive theoretical framework for analyzing the full spatiotemporal structure of both pulsating and snake solitary waves in the complex cubic-quintic Ginzburg-Landau equation. This includes elucidating the mechanism operative in creating these new classes of solitons in dissipative systems, as well as their absence in Hamiltonian and integrable systems where only stationary solitons are observed to occur.  The results obtained analytically using the
variational approximation for the snaking soliton is compared with
the numerical simulations of the CCQGLE using the RK4 method.

The specific theoretical modeling includes the use of a variational formulation and significantly generalized trial function for the solitary waves solutions. In addition, the resulting Euler-Lagrange equations are treated in an entirely different way by looking at their stable periodic solutions (or limit cycles) resulting from supercritical Hopf bifurcations. Oscillations of their trial function parameters on these limit cycles provide the pulsations of the amplitude, width, and phase of the solitons. The model also allows for detailed predictions regarding the other issue of central interest for the pulsating and snake solitons, viz. the effect of each of the system parameters on the amplitude, width, period, and stability of the solitary waves.

Also, given the generality of the theoretical framework developed in this paper, it provides a platform for the detailed modeling of chaotic, creeping and erupting solitary waves, which are focus of current work in this area.

\section{Appendix}
\begin{enumerate}
\item  For the pulsating  solitons, the nonlinear functions  $f_i$, $i=1,2,3 $ from  \eqref{4.5} are shown in Fig. \ref{f123}.
\begin{figure}[ht!]
\begin{center}
{\includegraphics[width=0.75\textwidth]{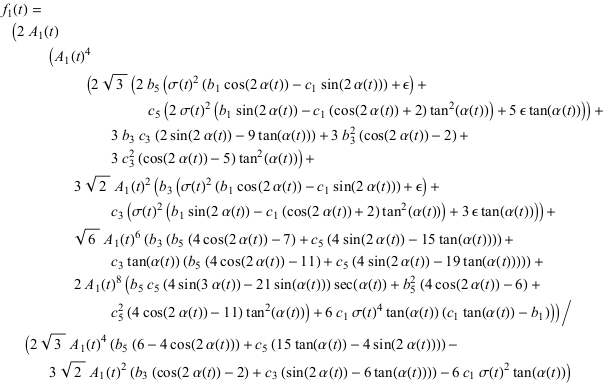}}\\
{\includegraphics[width=0.75\textwidth]{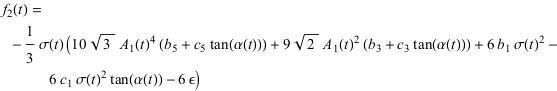}} \\
{\includegraphics[width=0.75\textwidth]{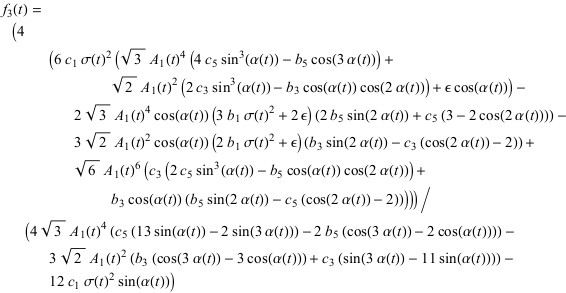}} 
\end{center}
\caption{Nonlinear functions  $f_i$, $i=1,2,3$ from \eqref{4.5}}\label{f123}
\end{figure}

\item  For the snaking solitons, the nonlinear functions  $f_i$, $i=4,5,6$  from \eqref{5.5} are shown in Fig. \ref{f456}.
\begin{figure}[ht!]
\begin{center}
{\includegraphics[width=0.75\textwidth]{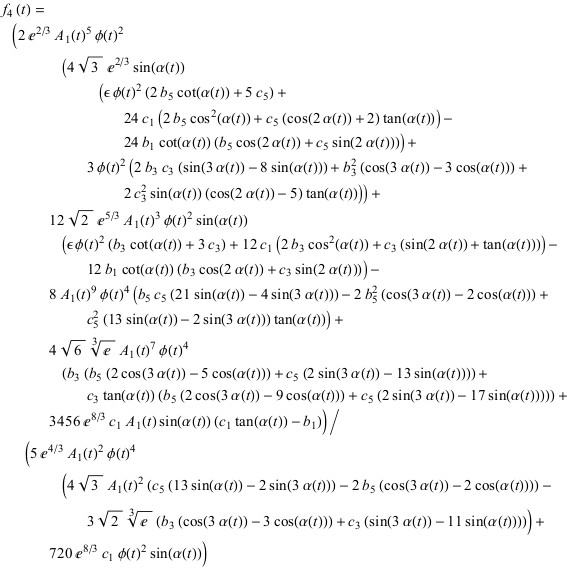}}\\
{\includegraphics[width=0.75\textwidth]{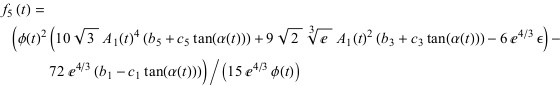}} \\
{\includegraphics[width=0.75\textwidth]{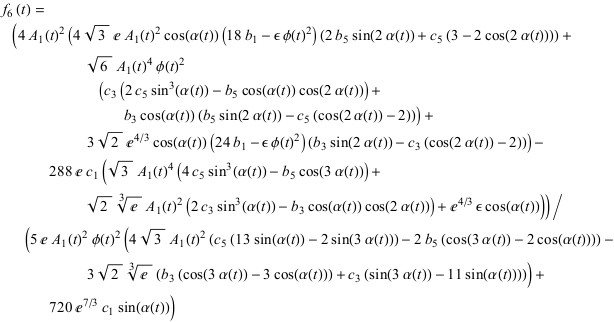}} 
\end{center}
\caption{Nonlinear functions  $f_i$, $i=4,5,6$ from \eqref{5.5}}\label{f456}
\end{figure}
\end{enumerate}

\section*{Acknowledgment}
We would like to acknowledge extremely insightful comments by David Kaup on the variational formulation, as well as on soliton perturbation theory. Helpful inputs were also provided by Jianke Yang, Roberto Camassa \cite{private}, and Harihar Khanal on the implementation of the RK4 method.

\bibliographystyle{plain}
\bibliography {bibliography}   

%%%%%%%%%%%%%%%%%%%%%%%%%%%%%%%%%%%%%%%%%%%%%%%%%%%%%%
\begin{figure}[!ht]
\begin{center}
{\includegraphics[width=.5\textwidth]{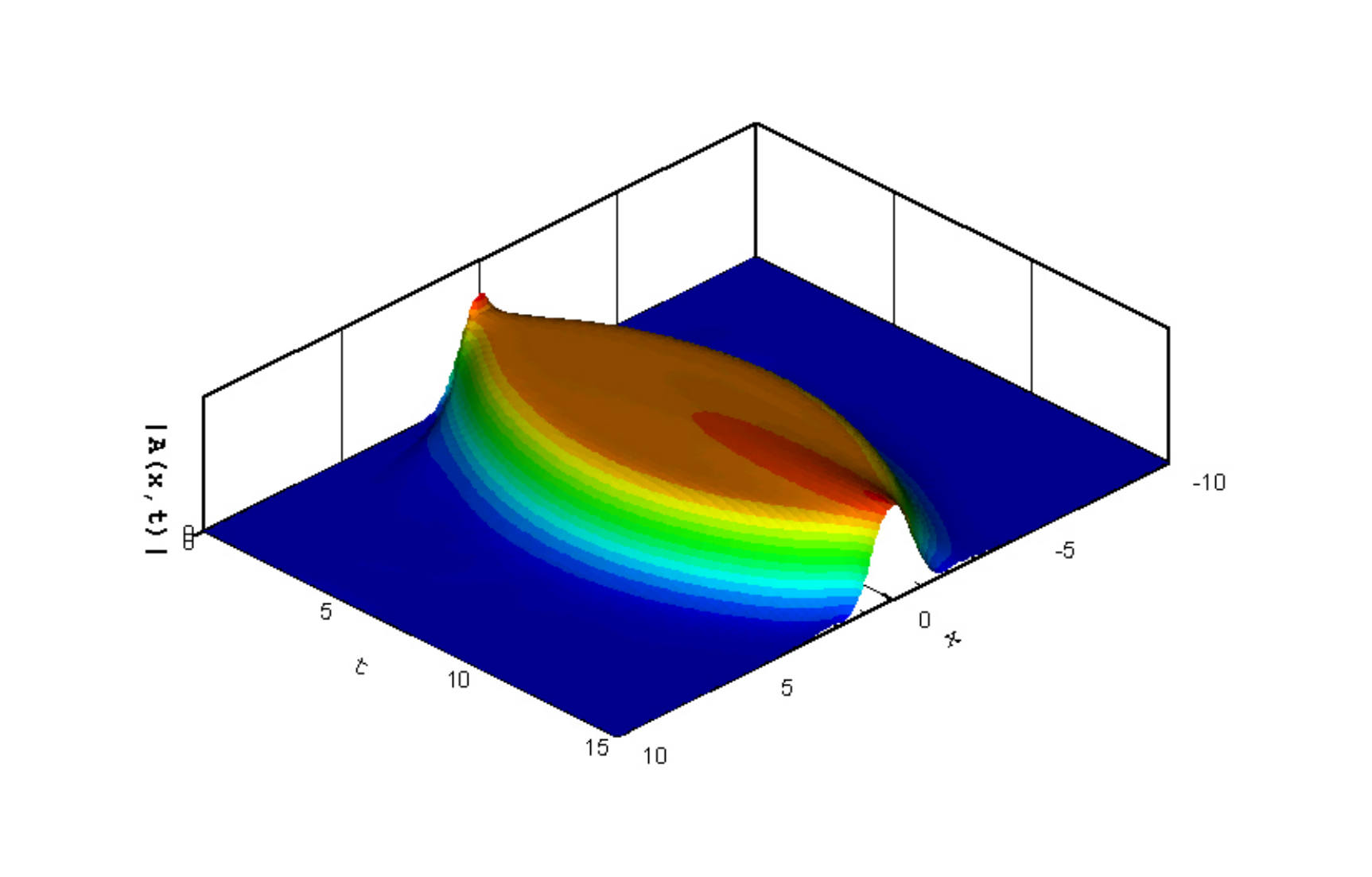}}\hspace{.1\textwidth}
{\includegraphics[width=.5\textwidth]{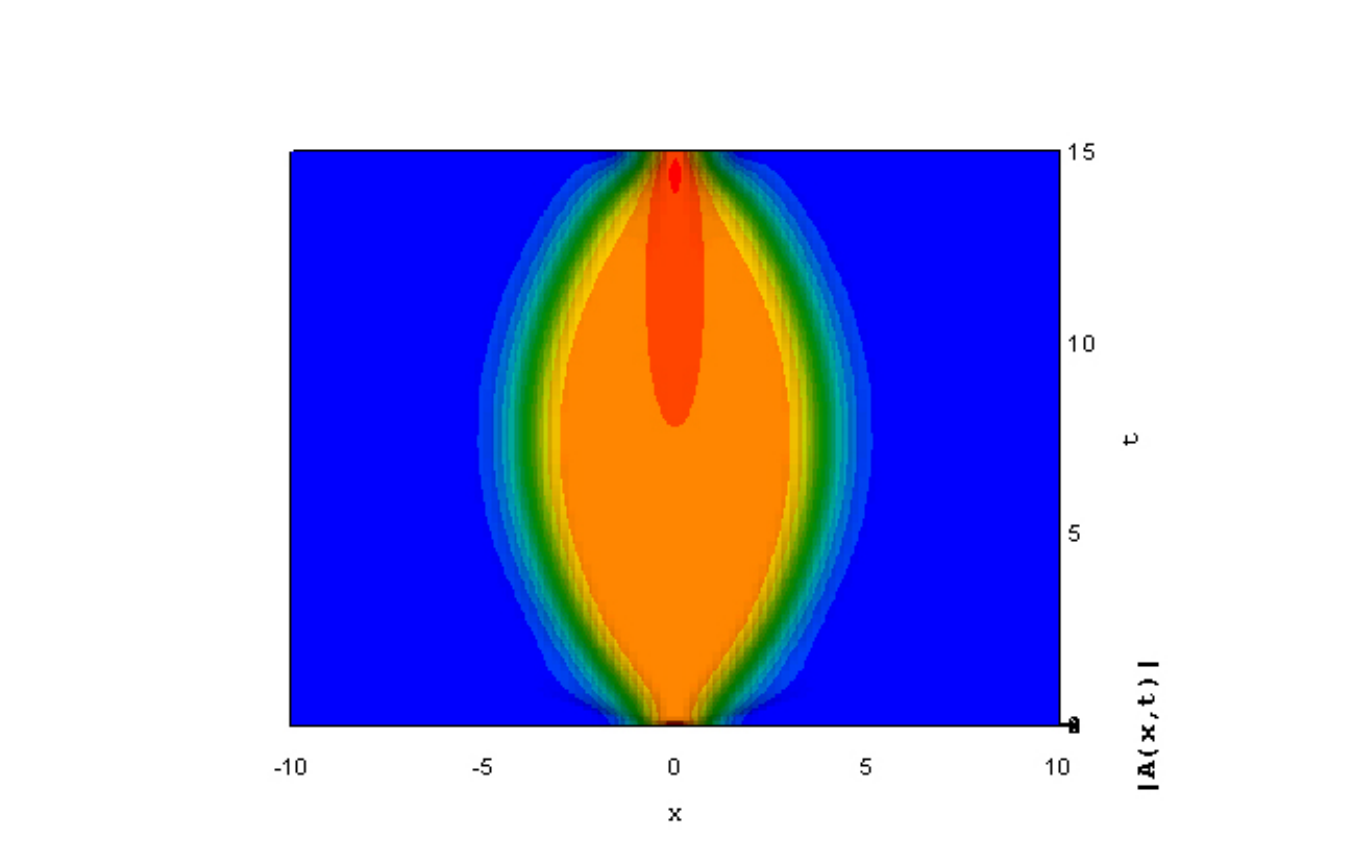}}\hspace{.1\textwidth}
{\includegraphics[width=.6\textwidth]{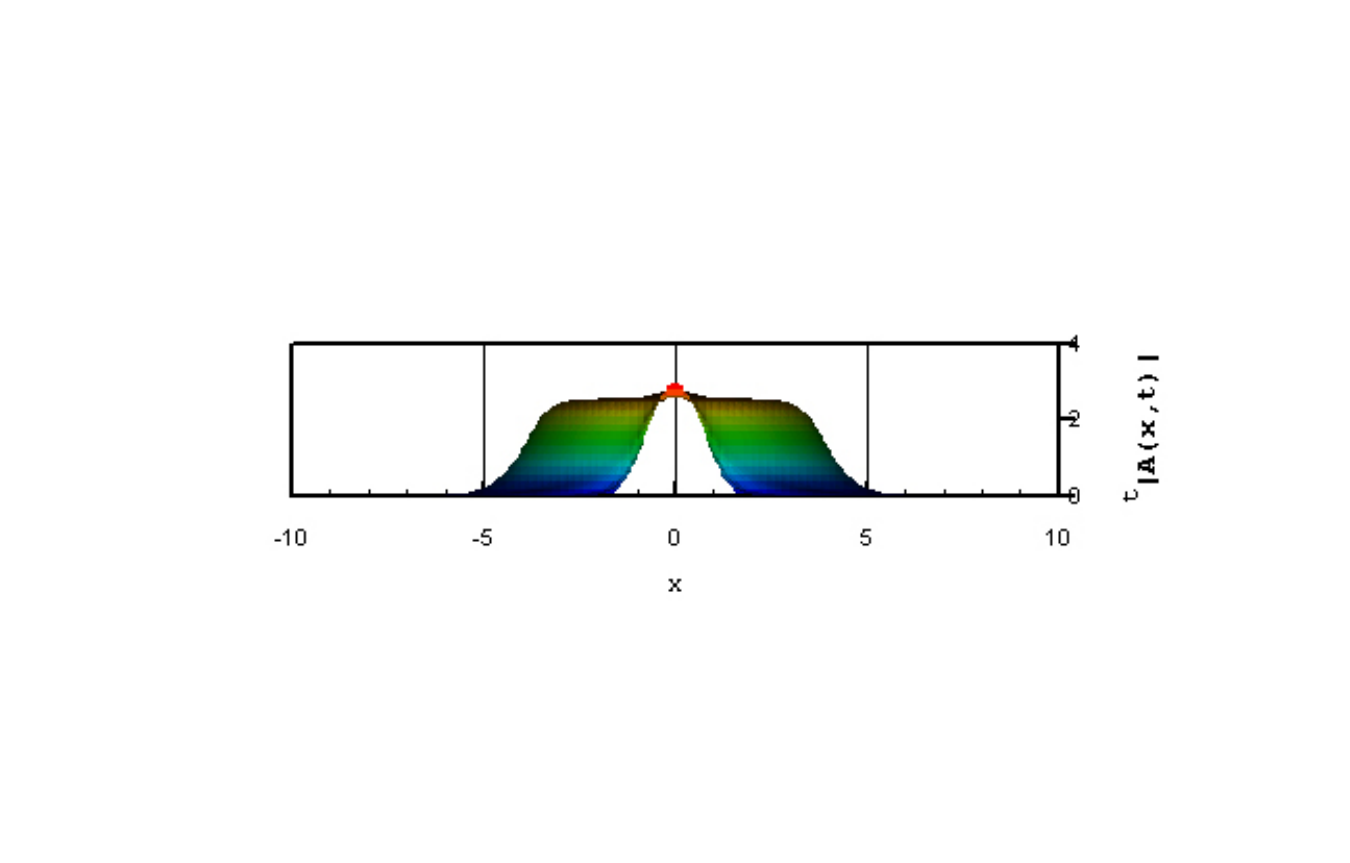}}\hspace{.1\textwidth}
\end{center}
\caption{Plain pulsating soliton for $b_3=-0.66$ and $\epsilon=-0.1$,  $b_1=0.08$, $b_5= 0.1$, $c_1=0.5$, $c_3=1$, $c_5=-0.1$} \label{Set1}
\end{figure}

\begin{figure}[!ht]
\begin{center}
{\includegraphics[width=.5\textwidth]{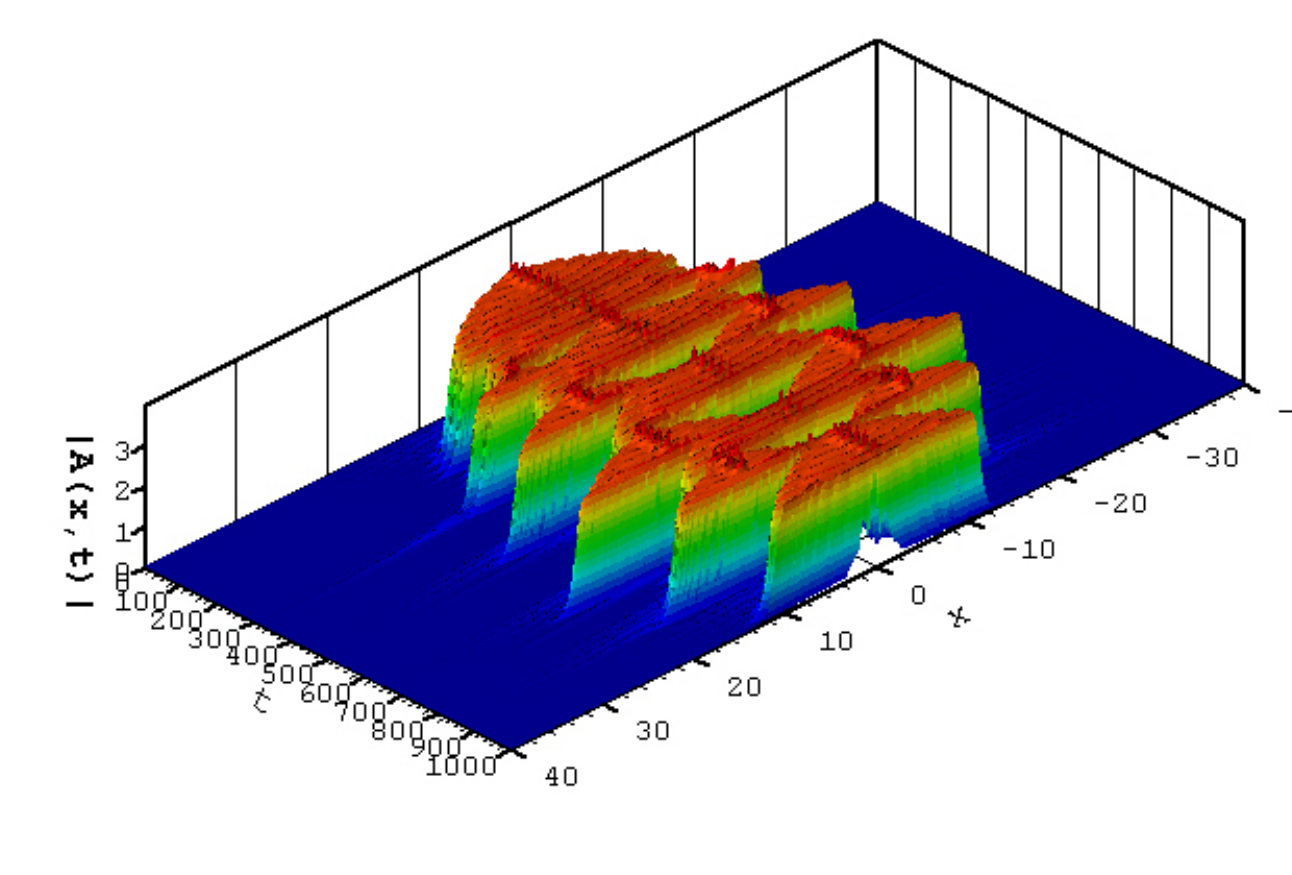}}\hspace{.1\textwidth}
{\includegraphics[width=.5\textwidth]{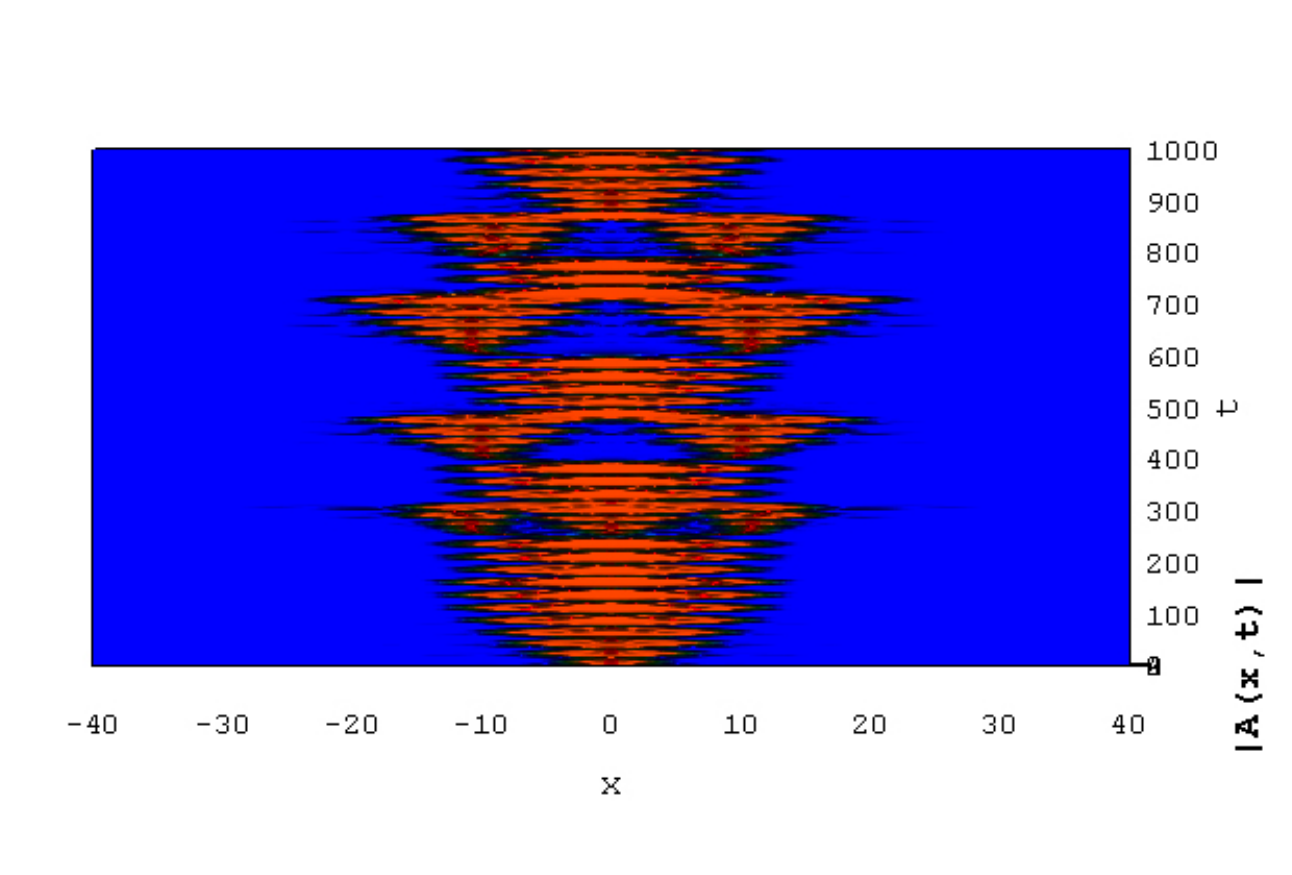}}\hspace{.1\textwidth}
{\includegraphics[width=.6\textwidth]{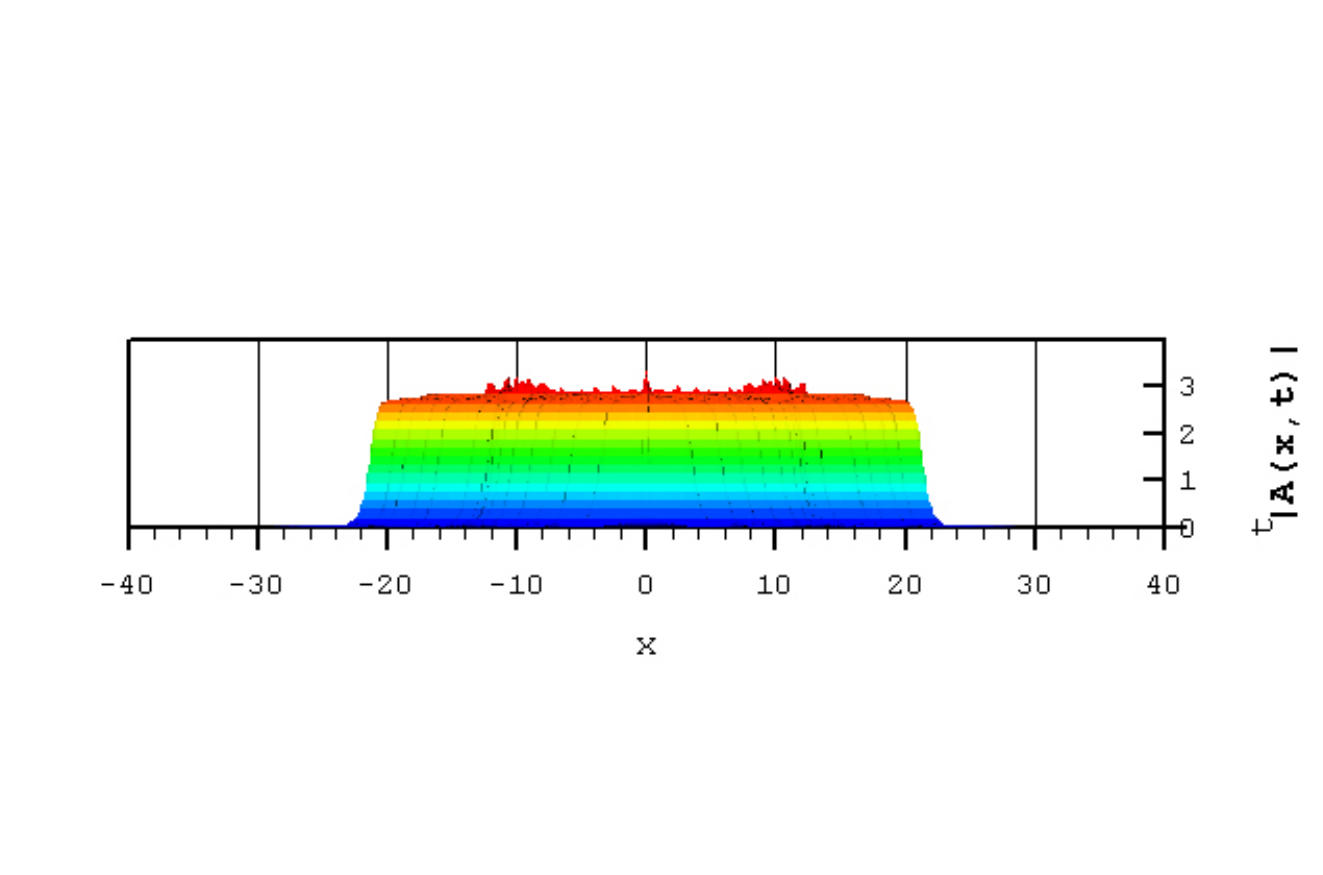}}\hspace{.1\textwidth}
\end{center}
\caption{Snake soliton  for $b_3=-0.66$ and $\epsilon=-0.1$,  $b_1=0.08$, $b_5= 0.11$, $c_1=0.5$, $c_3=1$, $c_5=-0.08$} \label{Set2}
\end{figure}

\begin{figure}[!ht]
\begin{center} 
{\includegraphics[width=.5\textwidth]{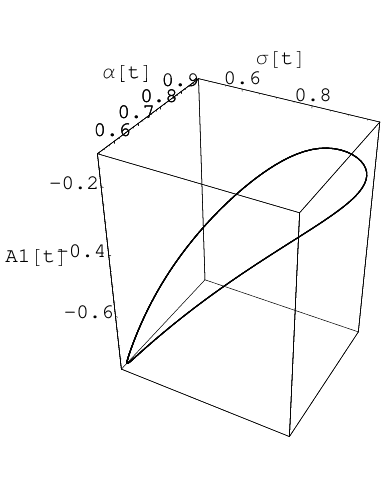}}\hspace{.1\textwidth}
{\includegraphics[width=.5\textwidth]{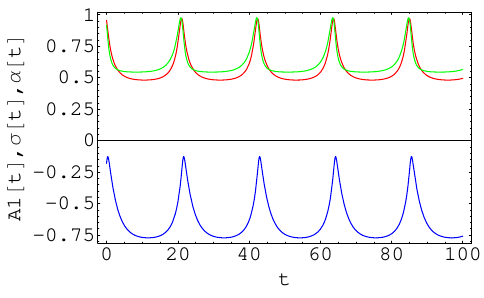}}\hspace{.1\textwidth}
{\includegraphics[width=.6\textwidth]{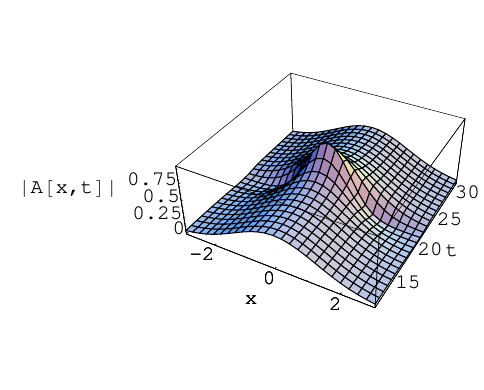}}\hspace{.1\textwidth}
\end{center}
\caption{The periodic orbit, periodic time series, and pulsating soliton  for $b_3=-0.1424$, $\epsilon=-0.345481$,  $b_1=0.08$, $b_5= 0.1$, $c_1=0.5$, $c_3=1$, $c_5=-0.1$}\label{Figure22}
\end{figure}

\begin{figure}[!ht]
					\begin{center} 
{\includegraphics[width=.5\textwidth]{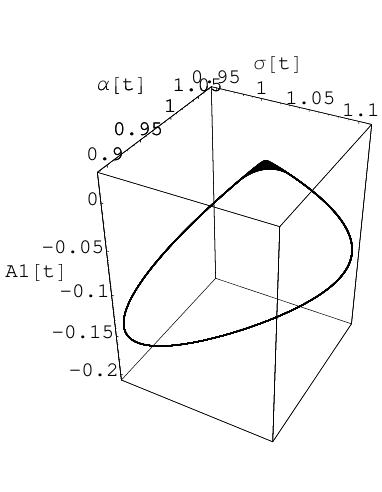}}\hspace{.1\textwidth}
{\includegraphics[width=.5\textwidth]{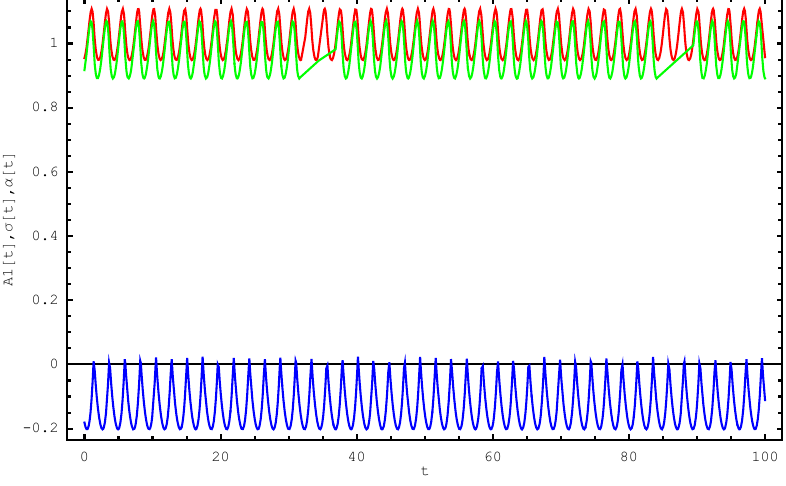}}\hspace{.1\textwidth}
{\includegraphics[width=.6\textwidth]{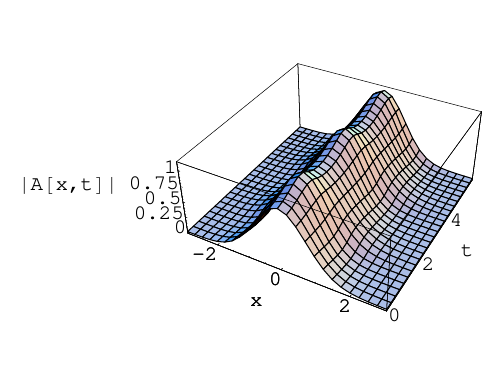}} \hspace{.1\textwidth}
					\end{center}
				\caption{The periodic orbit, periodic time series, and pulsating soliton  for $b_3=-0.2531943$, $\epsilon=-0.345481$, $b_1=0.08$, $b_5= 0.1$, $c_1=0.5$, $c_3=1$, $c_5=-0.1$}\label{Figure25}
\end{figure}

\begin{figure}[!ht]
					\begin{center} 
{\includegraphics[width=.5\textwidth]{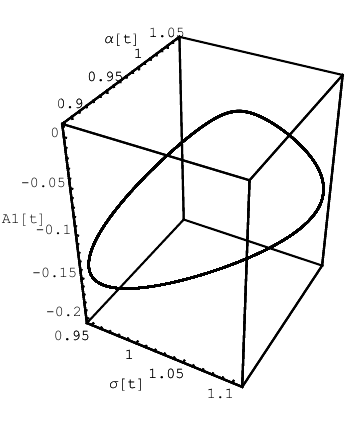}}\hspace{.1\textwidth}
{\includegraphics[width=.5\textwidth]{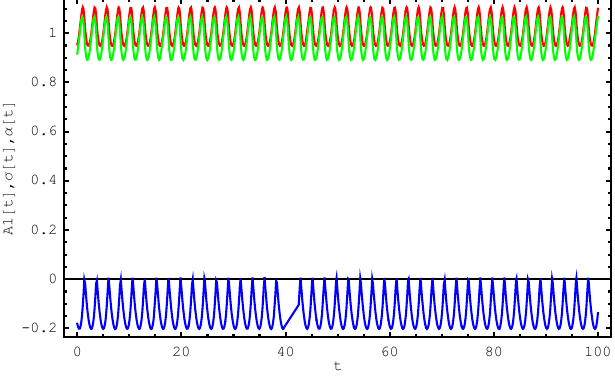}}\hspace{.1\textwidth}
{\includegraphics[width=.6\textwidth]{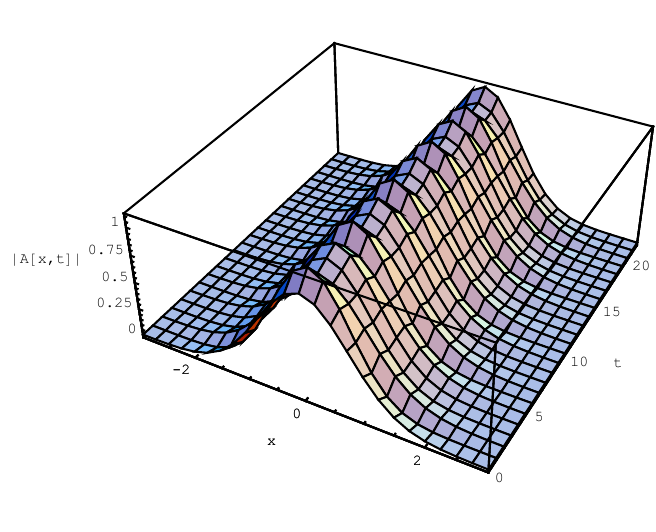}}\hspace{.1\textwidth}
					\end{center}
				\caption{The periodic orbit,  periodic time series, and pulsating soliton for $b_3=-0.2516$, $\epsilon=-0.345481$, $b_1=0.08$, $b_5= 0.1$, $c_1=0.5$, $c_3=1$, $c_5=-0.1$}\label{Figure27}
\end{figure}

\begin{figure}[ht!]
\begin{center}
\subfloat[Numerical simulations $b_1=0.2$]
{\includegraphics[width=.4\textwidth]{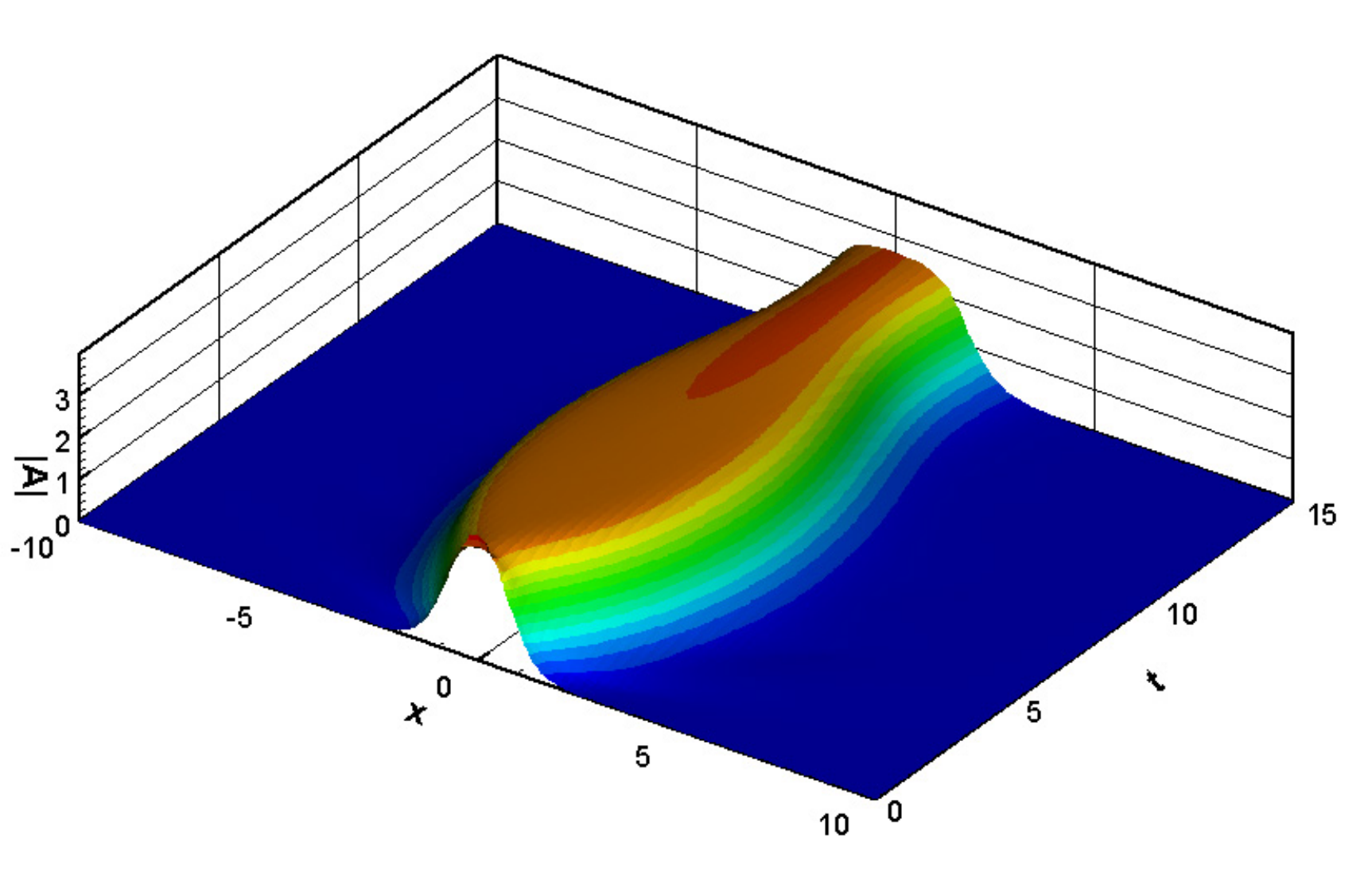}} \hspace{.1\textwidth}
\subfloat[Variational approximation $b_1=0.1$]
{\includegraphics[width=.4\textwidth]{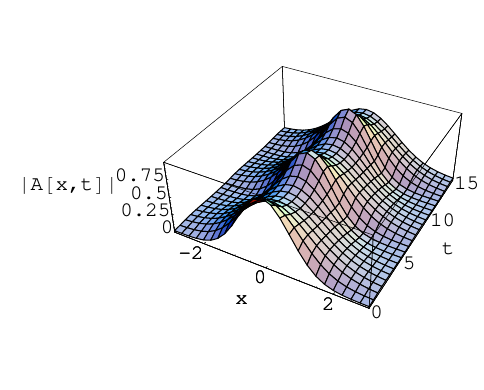}} \\
\subfloat[Numerical simulations $b_5=0.11$]
{\includegraphics[width=.4\textwidth]{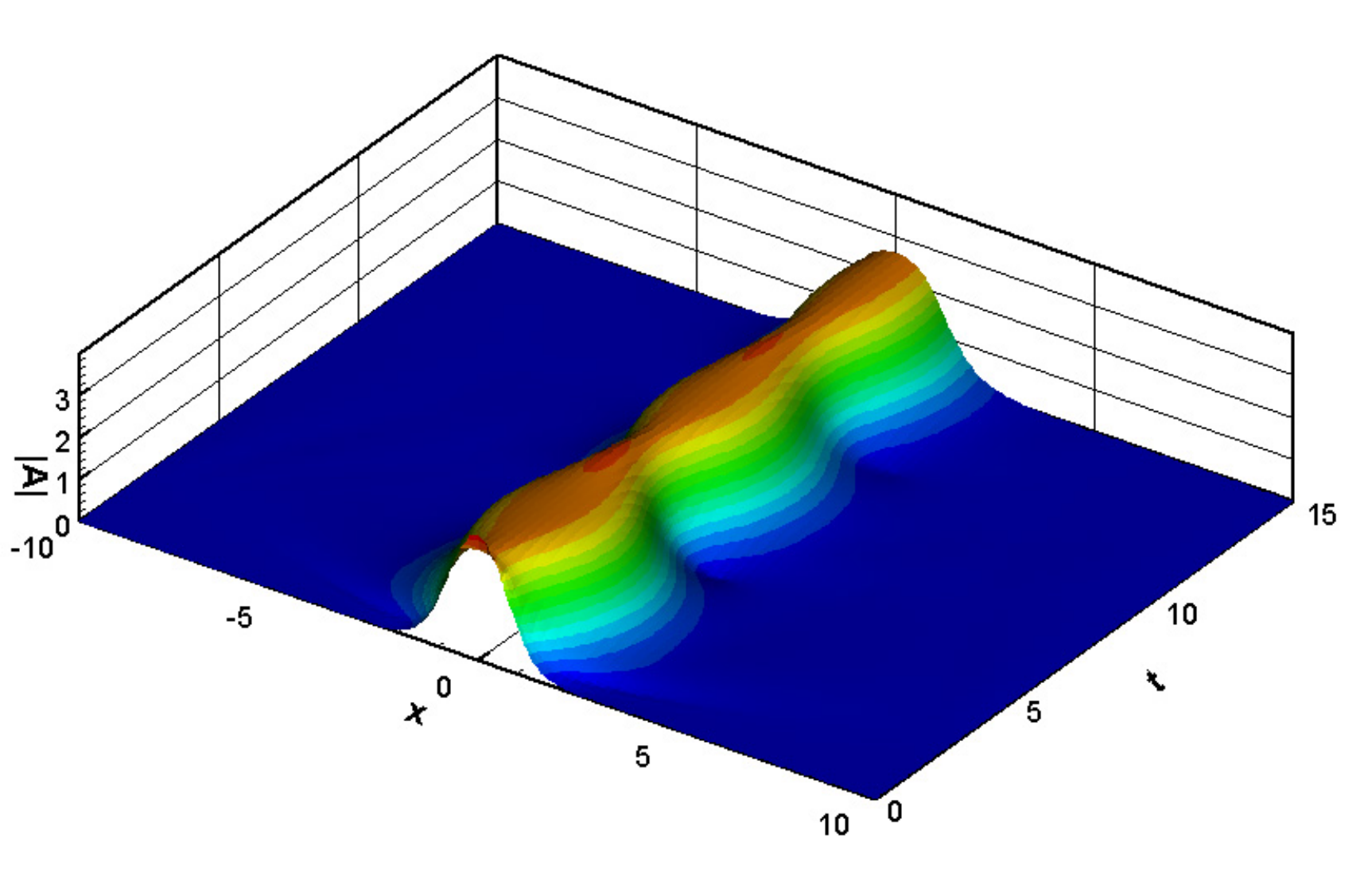}} \hspace{.1\textwidth}
\subfloat[Variational approximation $b_5=0.13$]
{\includegraphics[width=.4\textwidth]{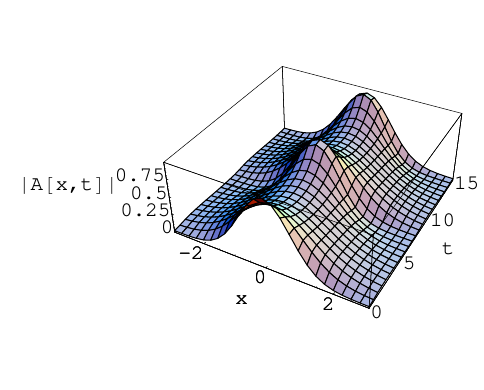}} \\
\subfloat[Numerical simulations $c_1=0.6$]
{\includegraphics[width=.4\textwidth]{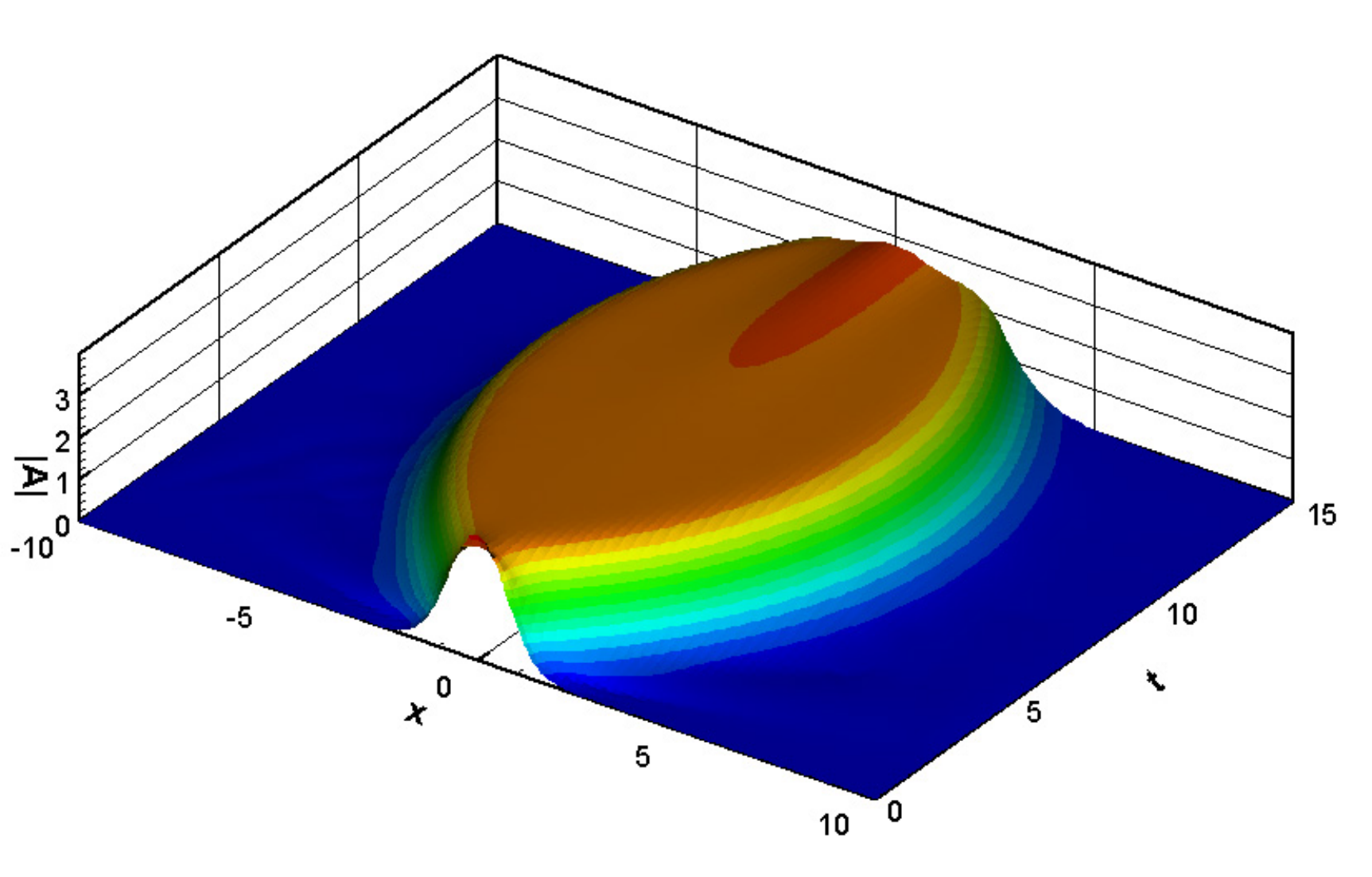}} \hspace{.1\textwidth}
\subfloat[Variational approximation $c_1=0.6$]
{\includegraphics[width=.4\textwidth]{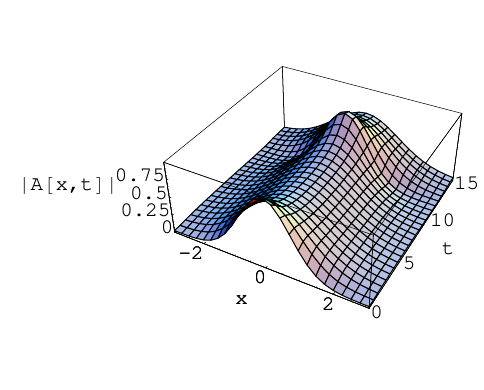}} \\
\end{center}
\caption{Predictions for the plane pulsating soliton cases (i)-(iii)}\label{plane10}
\end{figure}

\begin{figure}[ht!]
\begin{center}
\subfloat[Numerical simulations $c_3=1.05$]
{\includegraphics[width=.4\textwidth]{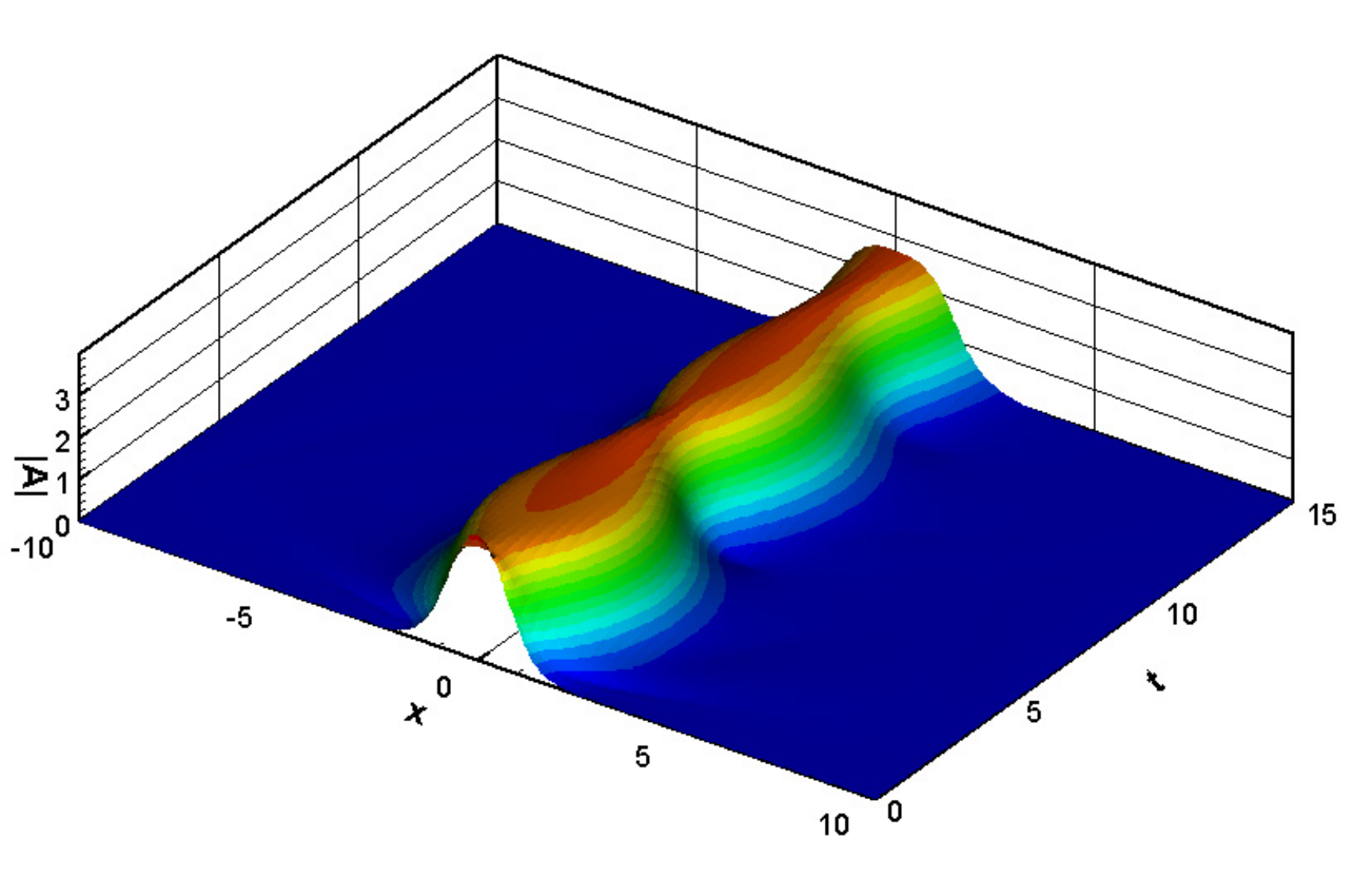}} \hspace{.1\textwidth}
\subfloat[Variational approximation $c_3=1.05$]
{\includegraphics[width=.4\textwidth]{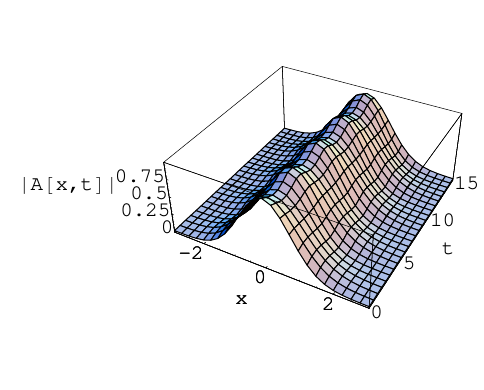}} \\
\subfloat[Numerical simulations $c_5=-0.075$]
{\includegraphics[width=.4\textwidth]{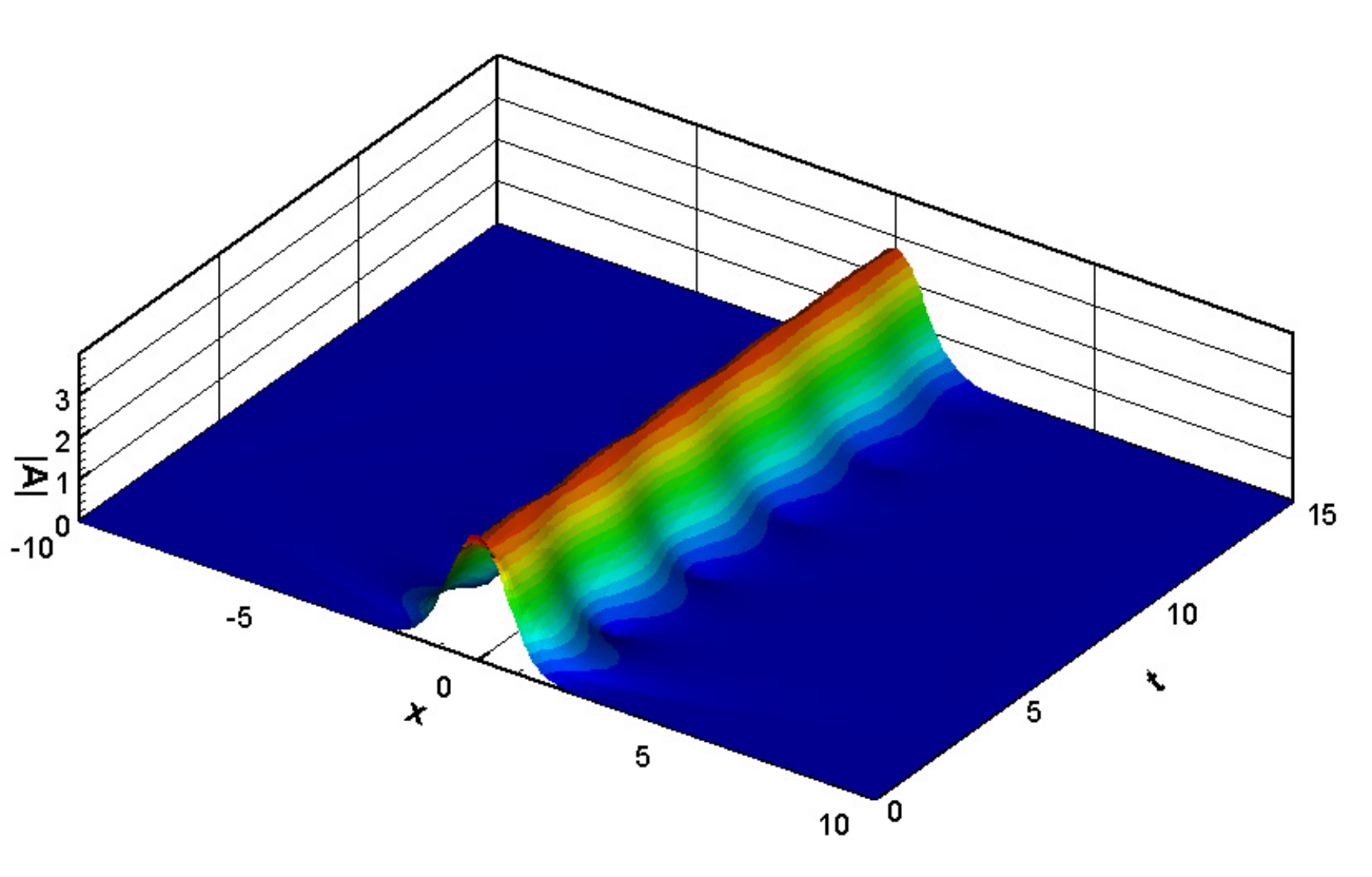}} \hspace{.1\textwidth}
\subfloat[Variational approximation $c_5=-0.08$]
{\includegraphics[width=.4\textwidth]{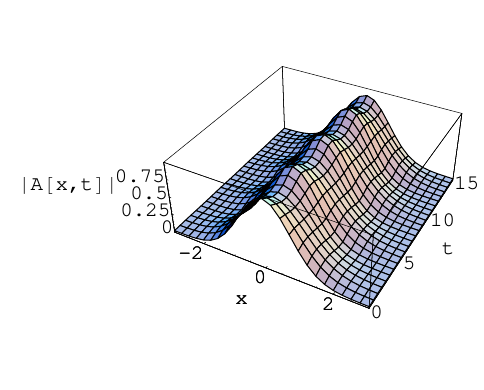}} \\
\subfloat[Numerical simulations $\epsilon=-0.08$]
{\includegraphics[width=.4\textwidth]{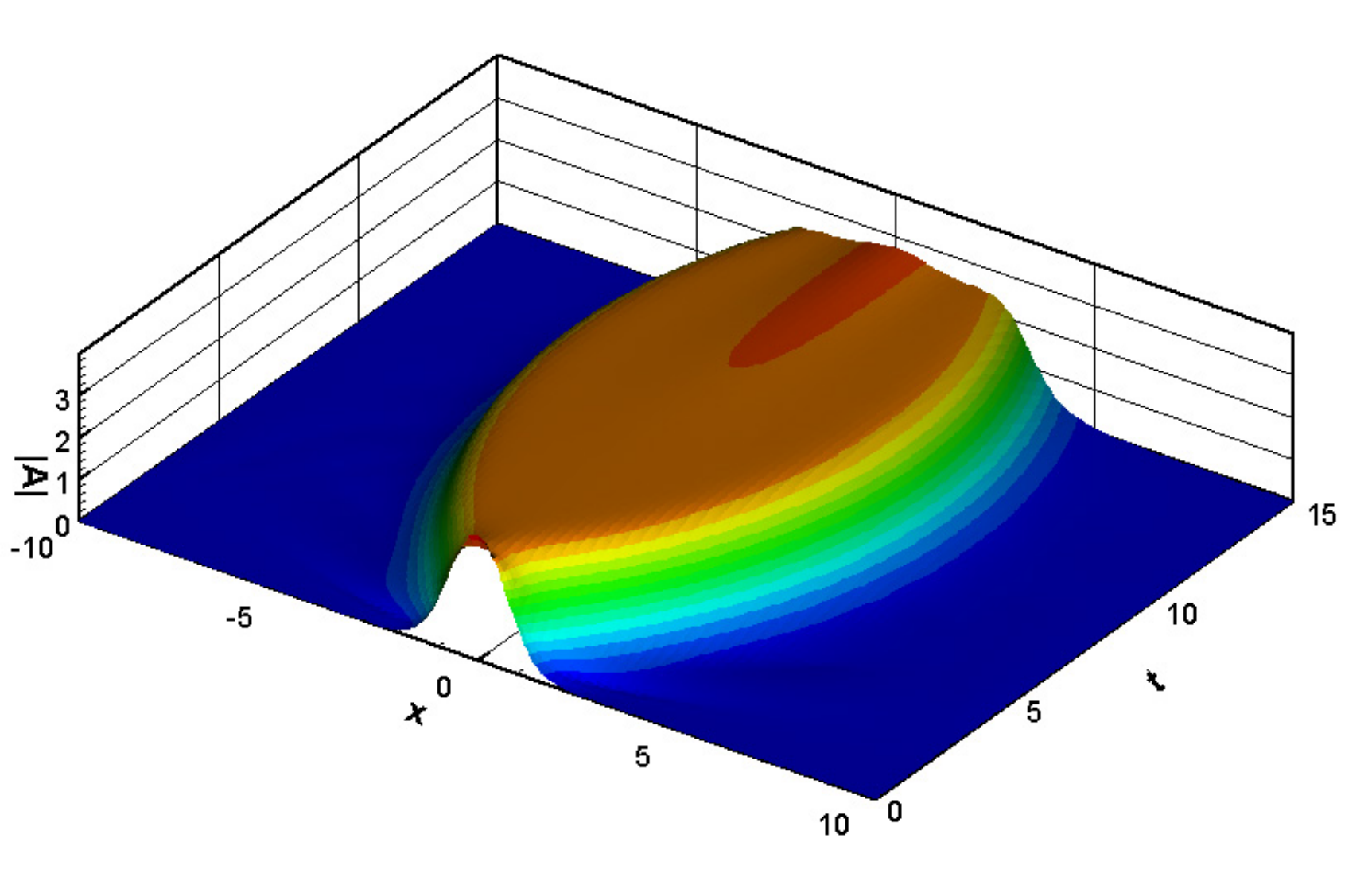}} \hspace{.1\textwidth}
\subfloat[Variational approximation $\epsilon=-0.06$]
{\includegraphics[width=.4\textwidth]{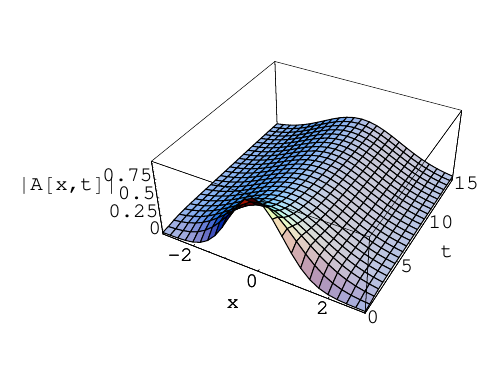}} \\
\end{center}
\caption{Predictions for the plane pulsating soliton cases (iv)-(vi)}\label{plane12}
\end{figure}

\begin{figure}[!ht]
\begin{center}
{\includegraphics[width=.5\textwidth]{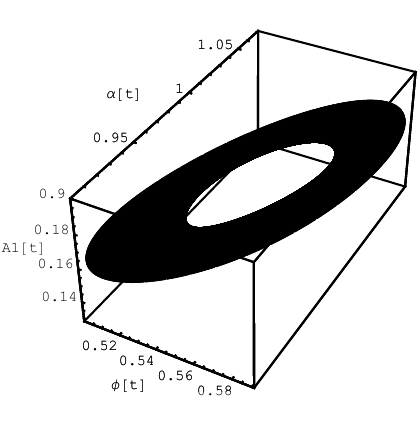}}\hspace{.1\textwidth}
{\includegraphics[width=.5\textwidth]{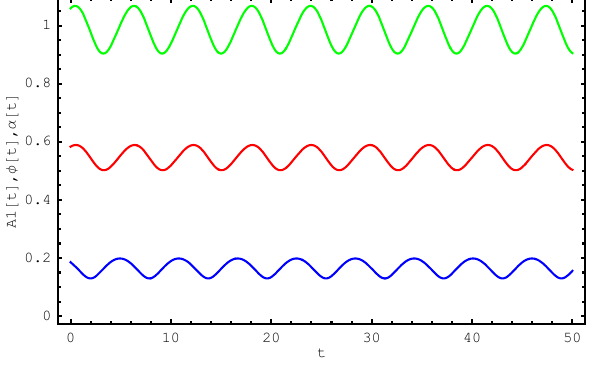}}\hspace{.1\textwidth}
{\includegraphics[width=.6\textwidth]{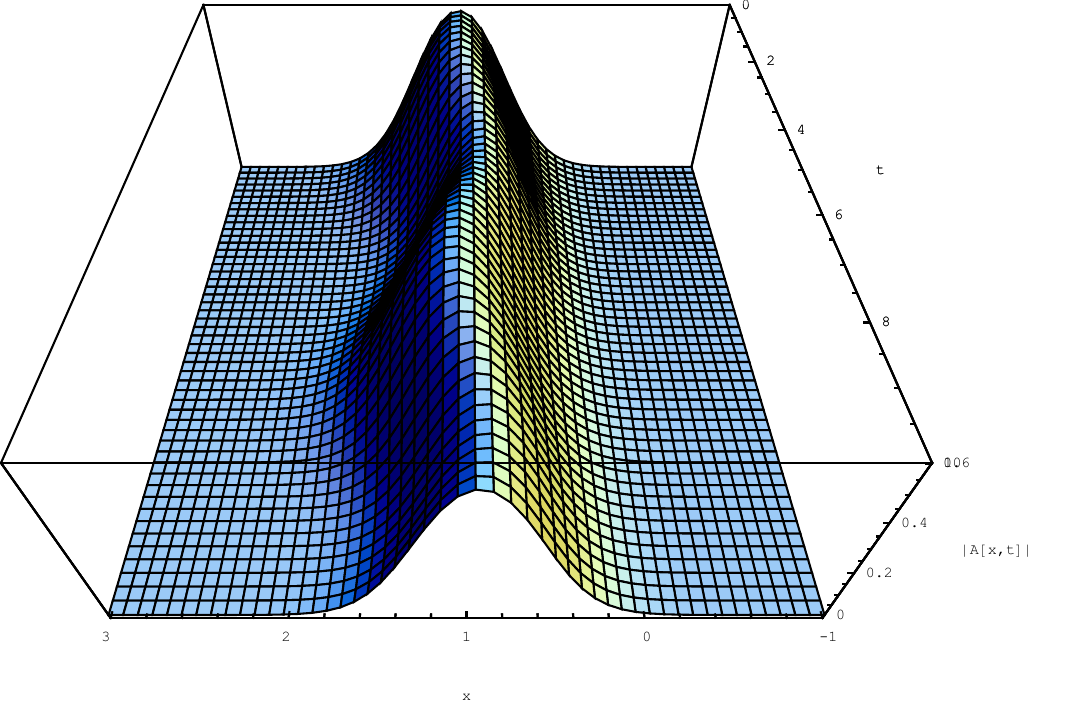}}\hspace{.1\textwidth}
\end{center}
\caption{Periodic orbit, periodic time series, and  snake soliton  for $b_3=-0.835$ and $\epsilon=-0.1$,  $b_1=0.08$, $b_5= 0.11$, $c_1=0.5$, $c_3=1$, $c_5=-0.08$} \label{Set6}
\end{figure}
\begin{figure}[!ht]
\begin{center}
{\includegraphics[width=.5\textwidth]{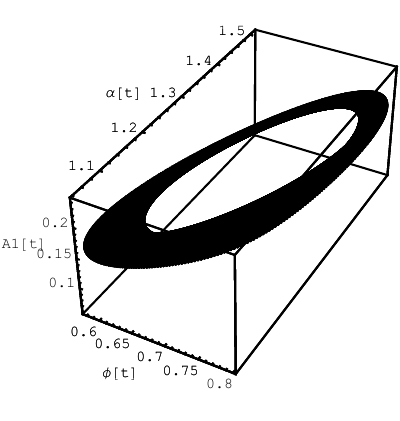}}\hspace{.1\textwidth}
{\includegraphics[width=.5\textwidth]{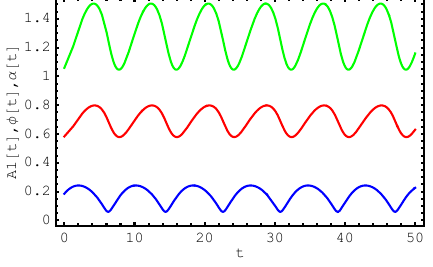}}\hspace{.1\textwidth}
{\includegraphics[width=.6\textwidth]{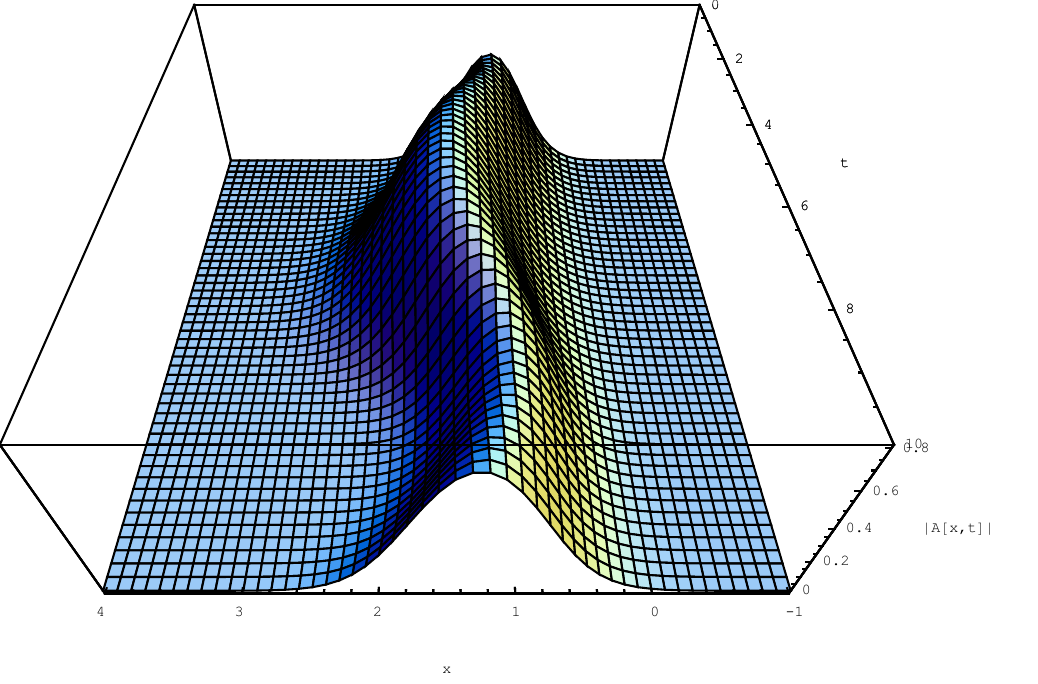}}\hspace{.1\textwidth}
\end{center}
\caption{Periodic orbit, periodic time series, and snake soliton  for $b_3=-0.61$ and $\epsilon=-0.1$,  $b_1=0.08$, $b_5= 0.11$, $c_1=0.5$, $c_3=1$, $c_5=-0.08$} \label{Set7}
\end{figure}
\begin{figure}[!ht]
\begin{center}
{\includegraphics[width=.5\textwidth]{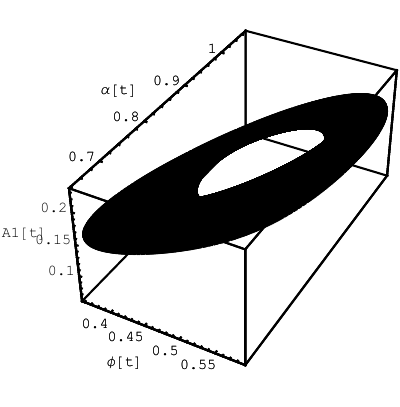}}\hspace{.1\textwidth}
{\includegraphics[width=.5\textwidth]{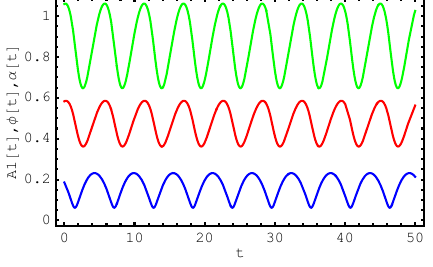}}\hspace{.1\textwidth}
{\includegraphics[width=.6\textwidth]{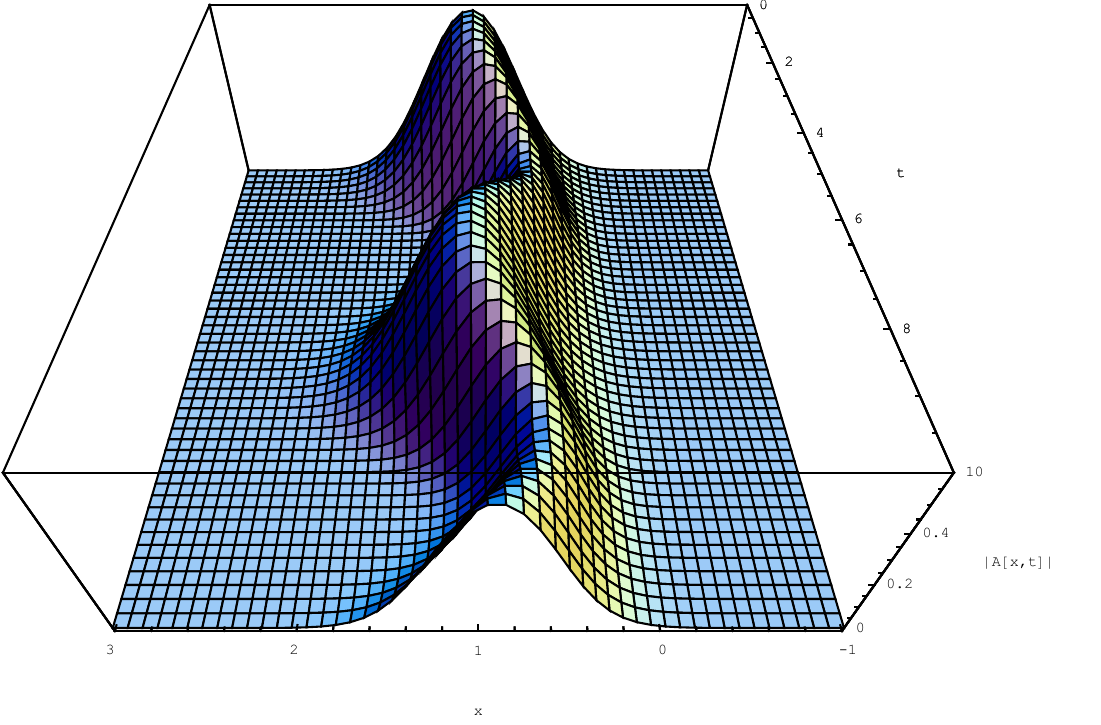}}\hspace{.1\textwidth}
\end{center}
\caption{Periodic orbit, periodic time series, and snake soliton  for $b_3=-0.66$ and $\epsilon=-0.08$,  $b_1=0.08$, $b_5= 0.11$, $c_1=0.5$, $c_3=1$, $c_5=-0.08$} \label{Set8}
\end{figure}

\begin{figure}[ht!]
\begin{center}
\subfloat[Numerical simulations $b_1=0.1$]
{\includegraphics[width=.4\textwidth]{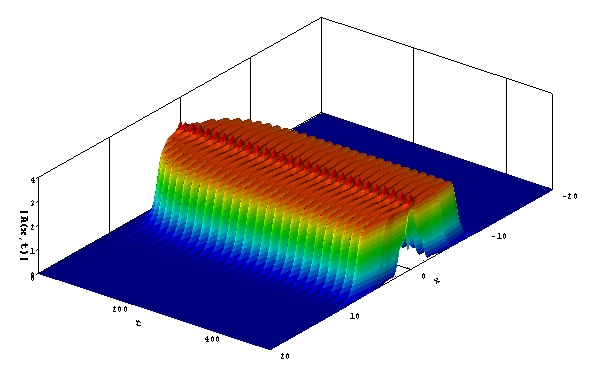}} \hspace{.1\textwidth}
\subfloat[Variational approximation $b_1=0.2$]
{\includegraphics[width=.4\textwidth]{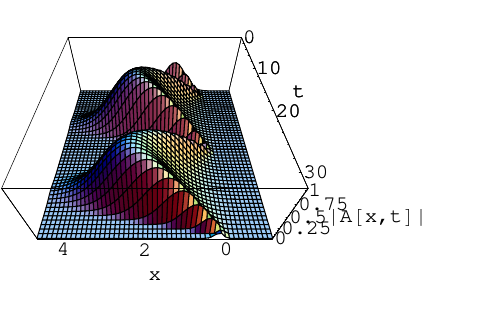}} \\
\subfloat[Numerical simulations $b_3=-0.8$]
{\includegraphics[width=.4\textwidth]{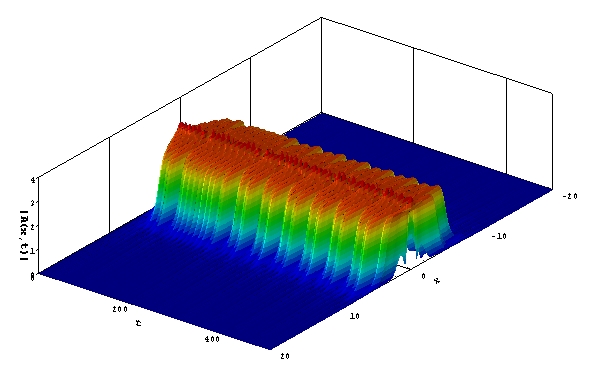}} \hspace{.1\textwidth}
\subfloat[Variational approximation $b_3=-0.6$]
{\includegraphics[width=.4\textwidth]{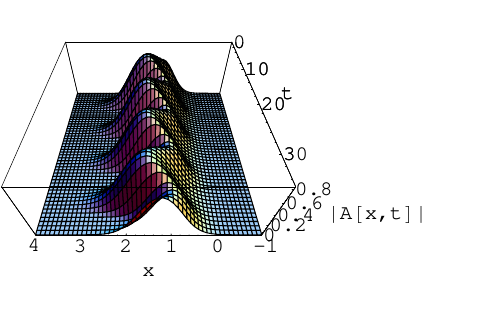}} \\
\subfloat[Numerical simulations $b_5=0.15$]
{\includegraphics[width=.4\textwidth]{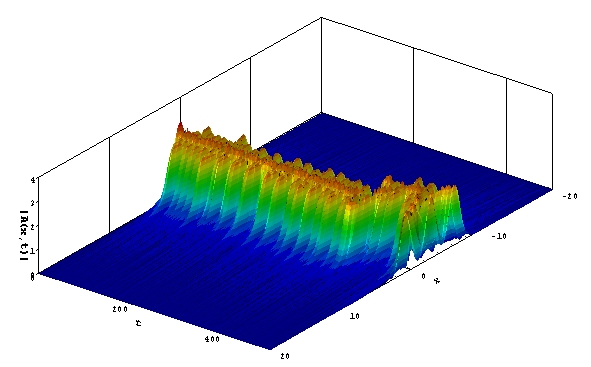}} \hspace{.1\textwidth}
\subfloat[Variational approximation $b_5=0.62$]
{\includegraphics[width=.4\textwidth]{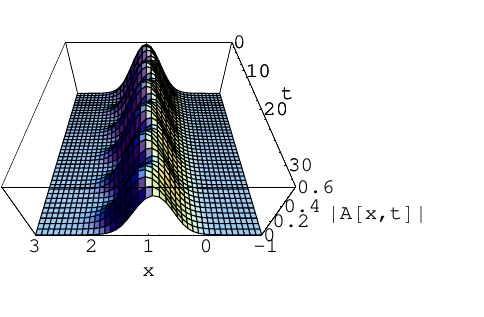}} \\
\subfloat[Numerical simulations $c_1=0.55$]
{\includegraphics[width=.4\textwidth]{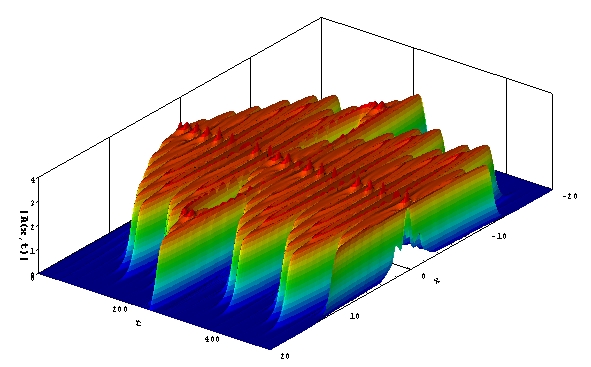}} \hspace{.1\textwidth}
\subfloat[Variational approximation $c_1=0.55$]
{\includegraphics[width=.4\textwidth]{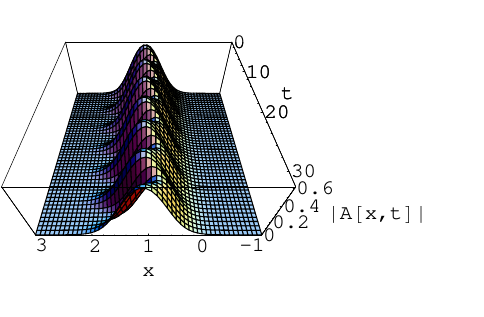}} \\
\end{center}
\caption{Predictions for snake solitons  cases (vii)-(x)}\label{snake12}
\end{figure}

\begin{figure}[ht!]
\begin{center}
\subfloat[Numerical simulations $c_3=1.135$]
{\includegraphics[width=.4\textwidth]{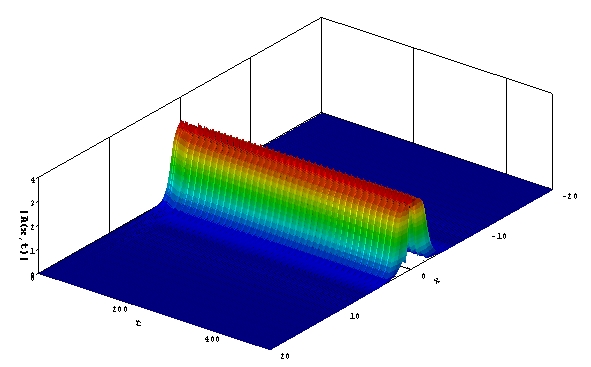}} \hspace{.1\textwidth}
\subfloat[Variational approximation $c_3=2$]
{\includegraphics[width=.4\textwidth]{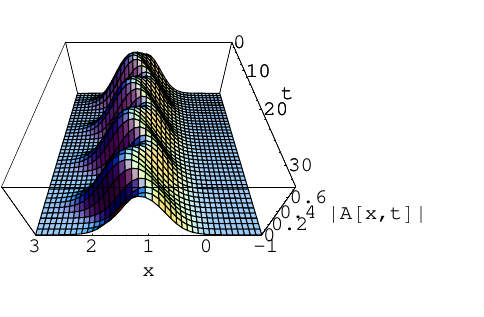}} \\
\subfloat[Numerical simulations $c_5=-0.06$]
{\includegraphics[width=.4\textwidth]{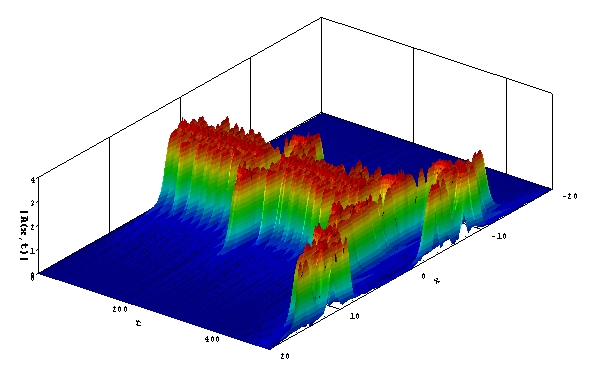}} \hspace{.1\textwidth}
\subfloat[Variational approximation $c_5=0.8$]
{\includegraphics[width=.4\textwidth]{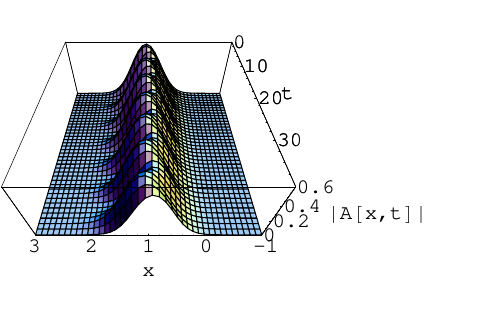}} \\
\subfloat[Numerical simulations $\epsilon=-0.05$]
{\includegraphics[width=.4\textwidth]{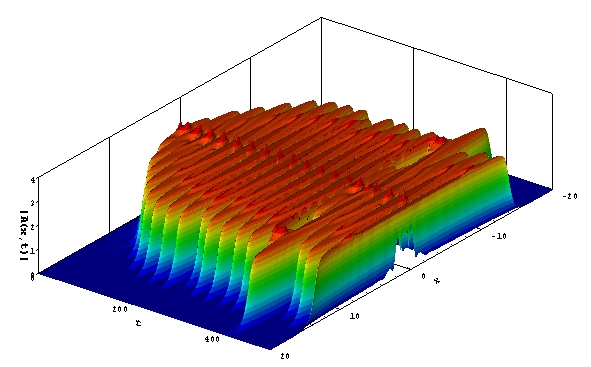}} \hspace{.1\textwidth}
\subfloat[Variational approximation $\epsilon=-0.08$]
{\includegraphics[width=.4\textwidth]{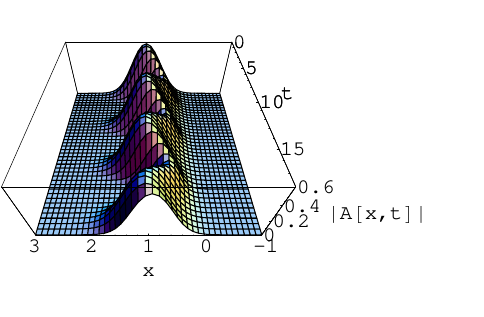}} \\
\end{center}
\caption{Predictions for snake solitons  cases (xi)-(xiii)}\label{snake13}
\end{figure}

\end{document}